\documentclass[journal]{inc/IEEEtran}
\usepackage{tikz}
\usetikzlibrary{decorations.pathmorphing, shapes}
\usetikzlibrary{calc}
\usetikzlibrary{arrows}
\usepackage{relsize}
\usepackage[americanvoltage]{circuitikz}
\usetikzlibrary{patterns}
\usetikzlibrary{decorations.markings}
\usetikzlibrary{arrows}

\ifCLASSINFOpdf
\else
\fi

\usepackage{amsmath}
\usepackage{amssymb}
\usepackage{mathtools}

\usepackage{booktabs}

\usetikzlibrary{calc}
\usepackage[americanvoltage]{circuitikz}
\usepackage{paralist}

\usepackage[caption=false]{subfig}

\usepackage{bm}
\newcommand{\bs}[1]{\bm{\mathrm{#1}}}

\newcommand{\vect}[1]{\vec{#1}}
\newcommand{\nhat}{\hat{{n}}}
\newcommand{\vecprime}[1]{\vect{#1}^{\,\prime}}

\newcommand{\matr}[1]{\bs{#1}}
\newcommand{\abs}[1]{\left \lvert #1 \right\rvert }
 
\newcommand{\junk}[1] {}

\def\XXint#1#2#3{{\setbox0=\hbox{$#1{#2#3}{\int}$}
\vcenter{\hbox{$#2#3$}}\kern-.5\wd0}}

\newcommand*\widebar[1]{%
  \hbox{%
    \vbox{%
      \hrule height 0.5pt 
      \kern0.3ex
      \hbox{%
        \kern-0.05em
        \ensuremath{#1}%
        \kern-0.05em
      }%
    }%
  }%
}

\renewcommand{\epsilon}{\varepsilon}

\newcommand{\divg}{\ensuremath{{\nabla}\cdot}}
\newcommand{\grad}{\ensuremath{\nabla}}

\newcommand{\figref}[1]{Fig.~\ref{#1}}
\newcommand{\tabref}[1]{Table~\ref{#1}}
\newcommand{\secref}[1]{Section~\ref{#1}}

\newcommand{\dyadic}[1]{\overline{\overline{\mathbf{#1}}}}

\usepackage{algorithm} 
\usepackage[noend]{algpseudocode} 

\makeatletter
\def\BState{\State\hskip-\ALG@thistlm}
\makeatother

\begin{document}

\title{A Complete Surface Integral Method for Broadband Modeling of 3D Interconnects in Stratified Media}

\author{Shashwat Sharma,~\IEEEmembership{Student~Member,~IEEE}, Utkarsh~R.~Patel,~\IEEEmembership{Student~Member,~IEEE}, Sean~V.~Hum,~\IEEEmembership{Senior~Member,~IEEE}, and 
        Piero~Triverio,~\IEEEmembership{Senior~Member,~IEEE}%
\thanks{Manuscript received ...; revised ...}%
\thanks{S.~Sharma, U.~R.~Patel, S. V.~Hum, and P.~Triverio are with the Edward S. Rogers Sr. Department of Electrical and Computer Engineering, University of Toronto, Toronto, M5S~3G4, Canada (emails: shash.sharma@mail.utoronto.ca, utkarsh.patel@mail.utoronto.ca, sean.hum@utoronto.ca, piero.triverio@utoronto.ca). }
\thanks{This work was supported by the Natural Sciences and Engineering Research 
	Council of Canada (Collaborative Research and Development Grants 
	program), by Advanced Micro Devices, and by CMC Microsystems.}

}

\markboth{IEEE Transactions on Components, Packaging and Manufacturing Technology}{IEEE Transactions on Components, Packaging and Manufacturing Technology}%

\maketitle
\begin{abstract}
	\boldmath
	A surface integral equation solver is proposed for fast and accurate simulation of interconnects embedded in stratified media. A novel technique for efficient computation of the multilayer Green's function is proposed. Using the Taylor expansion of Bessel functions, the computation of Sommerfeld integrals during the method of moments procedure is reduced to simple algebraic operations. To model skin effect in conductors, the single-source differential surface admittance operator is extended to conductors in stratified media. To handle large realistic structures, the adaptive integral method is developed for a multilayer environment in a generalized manner that poses no restrictions on layout of conductors, and requires no special grid refinement, unlike previous works. The proposed method is made robust over a wide frequency range with the augmented electric field integral equation. Realistic structures of different shapes and electrical sizes are successfully analyzed over a wide frequency range, and results are validated against a commercial finite element tool.
\end{abstract}

\section{Introduction}

Electromagnetic (EM) simulation tools are essential in the design of modern integrated circuits (ICs), which are becoming increasingly intricate.
Due to the high operating frequency and compact volume of most ICs, there is strong EM coupling and cross-talk between their constituent conductors.
Circuit designers require quantitative predictions of this coupling, and therefore would benefit greatly from fast and accurate EM modeling of IC components over a broad frequency range, typically from DC to tens of gigahertz.
Obtaining broadband EM models of on-chip interconnects is especially difficult because they exhibit strong field variations over such a wide frequency range, due to skin, proximity, and substrate effects. Furthermore, the increasing density and complexity of conductor layouts can lead to prohibitive computational costs.

Full-wave 3D characterization of interconnects requires solving for EM fields both inside (``interior problem'') and outside (``exterior problem'') conductors.
The interior problem is responsible for modeling skin and proximity effects.
The exterior problem captures coupling between conductors, as well as the effect of the surrounding medium.

Volumetric integral equation methods based on the method of moments (MoM) are well established for IC analysis~\cite{pfftmain}--\nocite{VolIE01, PEEC02}\cite{PEEC02}, particularly when coupled with acceleration techniques such as the adaptive integral method (AIM)~\cite{AIMbles} and the fast multipole method (FMM)~\cite{FMA}. However, a volumetric discretization is prohibitively expensive for large problems, particularly at high frequencies.
Surface-based approaches~\cite{gibson, aefie} can significantly reduce computational cost, but their widespread adoption has been limited by several open issues in both the exterior and interior problem. These challenges need to be addressed in order to robustly tackle realistic structures.

An open issue in solving the exterior problem is the accurate modeling of the stratified media in which interconnects are embedded. This is especially difficult for general configurations of conductors, where vias may traverse multiple dielectric layers.
An existing technique is the Poggio-Miller-Chang-Harrington-Wu-Tsai (PMCHWT) formulation~\cite{PMCHWT01}, which requires meshing of layer interfaces, leading to a large set of unknowns.
Furthermore, the PMCHWT formulation is not robust for material interfaces involving a large contrast in electrical properties.
An alternative is to use the multilayer Green's function (MGF), which does not require meshing of layer interfaces. However, analytical expressions do not exist for the MGF~\cite{MGF01}.
Instead, it is necessary to numerically compute semi-infinite Sommerfeld-type integrals for every combination of source and observation coordinates~\cite{SI_PE}, which is expensive. In both approaches, accounting for background stratification significantly increases computational cost.
Thus, there is a need for more robust and efficient techniques to model multilayered media.

A popular approach for computing the MGF more quickly is the discrete complex image method (DCIM)~\cite{DCIM01, DCIM03}, in which a fitting technique is used to approximate the MGF as a sum of exponential factors, called images, that correspond to spherical waves. For these factors, the Sommerfeld integral is known analytically. However, the computational cost associated with DCIM is proportional to the number of exponential terms needed. For realistic substrates, up to $10$-$15$ terms may be required, leading to a proportional increase in computation time compared to the homogeneous Green's function.

More recently, series expansions have been proposed for approximating the Green's function more efficiently~\cite{taylor1, konno}.
A Taylor expansion can be applied to the homogeneous Green's function to accelerate the partial element equivalent circuit method~\cite{taylor1}.
For layered media, a different approach can be taken~\cite{konno}, but it requires interpolation in addition to a series expansion. Additionally, numerical integration is required to tabulate interactions between source and observation pairs along two dimensions, which can be time-consuming.

Open issues also exist in efficiently solving the interior problem. Skin effect modeling in surface formulations is typically addressed using an analytically- or numerically-derived surface operator.
The purpose of this operator is to relate the tangential electric and magnetic fields on the surfaces of conductors.
Analytic operators based on the surface impedance boundary condition (SIBC)~\cite{YuferevIda_2009} are popular due to their simplicity. 
However, they generally require restrictive assumptions on geometry or frequency. For example, the SIBC is only accurate at high frequencies, where skin depth is significantly smaller than the conductor cross-section.
More rigorous analytic operators based on the differential surface admittance concept have been proposed~\cite{DSA01}--\nocite{DSA06}\cite{DSA07}, but are only applicable to canonical conductor geometries.

Numerical surface operators involve solving for additional integral equations for the interior problem, and allow accurate skin effect modeling for arbitrary conductor geometries~\cite{Chakraborty2006}--\nocite{Zhu2005,Tong2014,Chai2013,UTK_AWPL2017,Qian2007}\cite{UTK_EPEPS2017}. The generalized impedance boundary condition (GIBC)~\cite{Qian2007} and two-region surface integral equation methods~\cite{Tong2014, Chai2013} are two such approaches. However, these methods require solving for at least two sets of unknowns on the surfaces of conductors: an equivalent electric current density, and an equivalent magnetic current density. This necessitates generating additional integral operators, which increases computational cost. Furthermore, the additional operators require computing gradients of the complicated MGF, which further increases CPU time.

In this paper we propose a novel surface integral method for accelerated modeling of interconnects in stratified media. The proposed method addresses the issues previously discussed with the following new contributions:
\begin{itemize}
	\item Computation of the MGF is accelerated with a Taylor expansion of the Bessel function that appears in Sommerfeld integrands~\cite{epeps2018}, for which we provide a new mathematical and numerical analysis. Unlike the exponential images required with the DCIM, the proposed method only requires simple algebraic expressions during the MoM matrix fill.
	\item We propose a 2D fast Fourier transform (FFT)-based modification to AIM for multilayered media, whose computational cost is comparable to the conventional homogeneous case. Unlike previous works~\cite{AIM_MGF_2D}--\nocite{convcorr_AIM}\cite{okh_AIM_MGF_3D}, the proposed technique is applicable to any configuration of 3D conductors embedded in multiple dielectric layers.
	\item The open issue of the interior problem is addressed with the numerical differential surface admittance (DSA) operator~\cite{UTK_AWPL2017, UTK_EPEPS2017}, which we extend to multi-conductor systems embedded in stratified media. Unlike previous work on the DSA, the proposed approach is generalized to conductors traversing layer interfaces, by splitting such objects across layers and enforcing current continuity. The technique is thus suitable for any configuration of conductors embedded in an arbitrary stack-up of perfect or lossy dielectric layers. Unlike GIBC-based approaches, the DSA operator yields a single-source formulation that does not require gradients of the MGF.
\end{itemize}

This paper is organized as follows: \secref{sec:prob_statement} defines the problem. The MoM formulation for the interior and exterior problems is given in \secref{sec:MoM}. The proposed technique for accelerated MGF computation is described in \secref{sec:MGF}. Integration of the proposed approach with AIM is described in \secref{sec:AIM}. Finally, the complete solver is validated and tested on realistic structures, and the results are presented and discussed in \secref{sec:results}. The work is summarized and concluded in \secref{sec:concl}.

\section{Problem Statement} \label{sec:prob_statement}
Throughout this paper, we consider a general structure consisting of $N_c$ conductors with an arbitrary geometry, embedded in $N_d$ perfect or lossy dielectric layers with stratification along the $z$ axis. Conductors are excited by any number of user-defined ports. The goal is to solve Maxwell's equations in integral form with a surface-based triangle mesh on the conductors, and efficiently extract the scattering ($S$) matrix.
Conductors that traverse multiple dielectric layers are split at the layer interfaces, so that each sub-conductor is completely embedded in a single dielectric layer. The resulting sub-conductors are in contact with each other, and it is assumed that their common faces have a conformal mesh.

\subsection{Notation}
Throughout this work, we consider time-harmonic fields. A time dependence of $\mathrm{e}^{j\omega t}$ is assumed and suppressed. Field quantities are written with an overhead arrow, for example $\vect{a}\left(\vect{r}\right)$. Primed coordinates represent source points, while unprimed coordinates represent field observation points. Matrices and column vectors are written in bold letters, such as $\matr{A}$, while dyadic quantities are written with a double bar overhead, as in $\dyadic{B}$. Spectral domain quantities in terms of spatial frequency are denoted with an overhead tilde, such as $\widetilde{G}$.

\section{MoM System Formulation} \label{sec:MoM}
Modeling lossy conductors requires solving an interior and an exterior integral equation system. We employ the 3D numerical DSA operator~\cite{UTK_AWPL2017, UTK_EPEPS2017} to model the interior problem, and the A-EFIE \cite{aefie} for the exterior problem.

\subsection{Problem Geometry}

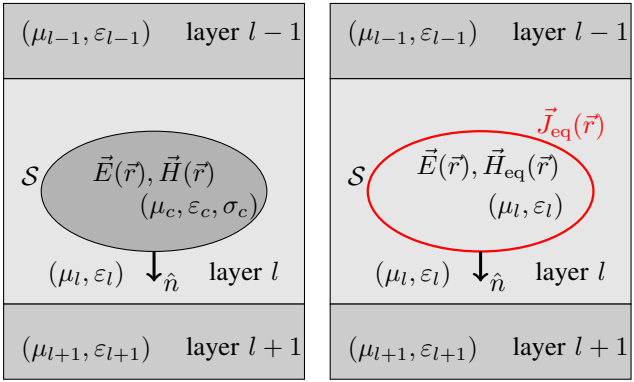
\begin{figure}[t]
	\centering
	\begin{tikzpicture}[scale = 1]
\draw[black, fill = black!10] (-2,-1.5) rectangle (2,1.5);
\draw[black, fill = black!20] (-2,-2.5) rectangle (2,-1.5);
\draw[black, fill = black!20] (-2,1.5) rectangle (2,2.5);
\draw[black, fill = black!30] (0,0) ellipse (1.5cm and 0.8cm);
\draw[black, ->, line width = 0.4mm] (0, -0.8) -- (0, -1.2);
\node at (0, -1.2) [right] {${\nhat}$};
\node at (0.6,-0.2) {$({\mu_c}, {\varepsilon_c}, \sigma_c)$};
\node at (0.0,0.6) [below] {$\vect{E}(\vect{r}), \vect{H}(\vect{r})$};
\node at (-1.4,0.2) [left]{${\cal S}$};
\node at (-0.9, -1.1) {$(\mu_l,\varepsilon_l)$};
\node at (-0.9, -2.1) {$(\mu_{l+1},\varepsilon_{l+1})$};
\node at (-0.9, 2.1) {$(\mu_{l-1},\varepsilon_{l-1})$};
\node at (1.2, -1.1) {layer $l$};
\node at (1.2, -2.1) {layer $l+1$};
\node at (1.2, 2.1) {layer $l-1$};
\end{tikzpicture}
	\begin{tikzpicture}[scale = 1]
\draw[black, fill = black!10] (-2,-1.5) rectangle (2,1.5);
\draw[black, fill = black!20] (-2,-2.5) rectangle (2,-1.5);
\draw[black, fill = black!20] (-2,1.5) rectangle (2,2.5);
\draw[red, fill = black!10, line width = 0.3mm] (0,0) ellipse (1.5cm and 0.8cm);
\draw[black, ->, line width = 0.4mm] (0, -0.8) -- (0, -1.2);
\node at (0, -1.2) [right] {${\nhat}$};
\node at (0.6,-0.2) {$(\mu_l, \varepsilon_l)$};
\node at (0.1,0.7) [below] {$\vect{E}(\vect{r}), {\vect{H}_\mathrm{eq}}(\vect{r})$};
\node at (-1.4,0.2) [left]{${\cal S}$};
\node at (-0.9, -1.1) {$(\mu_l,\varepsilon_l)$};
\node at (-0.9, -2.1) {$(\mu_{l+1},\varepsilon_{l+1})$};
\node at (-0.9, 2.1) {$(\mu_{l-1},\varepsilon_{l-1})$};
\node at (1.2, -1.1) {layer $l$};
\node at (1.2, -2.1) {layer $l+1$};
\node at (1.2, 2.1) {layer $l-1$};
\node at (1.2, 0.9) {${\color{red} \vect{J}_\mathrm{eq}(\vect{r})}$};
\end{tikzpicture}
	\caption{Left panel (original configuration): cross-section of a sample 3D conductor inside a stratified medium. Right panel (equivalent configuration): the original conductor is replaced by the surrounding medium, and an equivalent current density, shown in red, is introduced on ${\cal S}$.}
	\label{fig:sample}
\end{figure}

Consider a single conductor, with conductivity $\sigma_c$, permeability $\mu_c$, and permittivity $\varepsilon_c$, fully embedded in layer $l$ of the substrate.
Treatment for conductors in contact will be discussed in \secref{sec:continuity}.
\figref{fig:sample} shows the cross-section of the sample conductor in the stratified medium.
The conductor's surface is denoted by ${\cal S}$, and $\nhat$ is its outward unit normal vector.
A triangular surface mesh with $N_e$ edges is generated for ${\cal S}$.

%
\subsection{Interior Problem via DSA} \label{sec:interior}

The tangential electric and magnetic fields on ${\cal S}$ are expanded with RWG basis functions~\cite{Rao1982} normalized by edge length,
\begin{align}
	\nhat \times \vect{E}(\vect{r}) &= \sum_{n=1}^{N_e} E_n \vect{f}_{n}(\vect{r}) \label{eq:E1}\\
	\nhat \times \vect{H}(\vect{r}) &= \sum_{n=1}^{N_e} H_n \vect{f}_{n}(\vect{r}) \label{eq:H1}\,.
\end{align}
The coefficients $E_n$ and $H_n$ are stored in column vectors $\matr{E}$ and $\matr{H}$, respectively.

%
The tangential electric and magnetic fields on ${\cal S}$ are related by the Stratton-Chu formulation~\cite{Gibson2014}, which can be discretized using the standard MoM procedure to get the matrix equation,
\begin{align}
	{\matr{K}}_{c} {\matr{H}} + j\omega\epsilon_c{\matr{L}}_{c} \matr{E} &= 0 \label{eq:MFIEi}\,,
\end{align}
where $\eta_c = \sqrt{\mu_c/\left(\epsilon_c - j\sigma_c/\omega\right)}$ is the wave impedance in the conductive medium. Terms $\mu_c$, $\epsilon_c$ and $\sigma_c$ are the permeability, permittivity and conductivity of the conductor. Equation~\eqref{eq:MFIEi} is a discretization of the magnetic field integral equation (MFIE), and subscript $c$ indicates that the homogeneous Green's function with material properties of the conductor should be used.
Rotated RWG testing functions are used in the MoM procedure, and entries for operators $\matr{L}_c$ and $\matr{K}_c$ can be found in literature~\cite{Gibson2014}.
The resulting operators are well-conditioned due to the large wave number in conductive media.
However, due to the sharp decay of the Green's function involving large wave numbers, specialized numerical integration techniques are required when forming $\matr{L}_c$ and $\matr{K}_c$. To this end, we employ the line integral technique proposed by Qian and Chew~\cite{GIBC}.

We can now express the magnetic field in terms of the electric field as
\begin{equation}
	\matr{H} = \underbrace{-\dfrac{1}{j\omega\epsilon_c}\left[ \matr{K}_{c} \right]^{-1} \matr{L}_{c}}_{\matr{Y}_{\mathrm{in}}} \matr{E}\,,\label{eq:Yin}
\end{equation}
where $\matr{Y}_{\mathrm{in}}$ is the surface admittance operator associated with the interior medium.
Although computing $\matr{Y}_{\mathrm{in}}$ requires the factorization of a full matrix, the size of $\matr{K}_{c}$ is typically small compared to the overall problem. If necessary, large conductors can be split up into smaller segments.

Next, the equivalence principle \cite{Bal89} is applied to replace the conductor by its surrounding medium, and to introduce an equivalent electric current density $\vect{J}_\mathrm{eq}(\vect{r})$ on $\mathcal{S}$, as shown in Fig.~\ref{fig:sample}.
The equivalent current is also expanded with edge-normalized RWG functions,
\begin{equation}
	\vect{J}_\mathrm{eq}(\vect{r}) = \sum_{n=1}^{N_e} J_n \vect{f}_{n}(\vect{r})\,. \label{eq:Jdisc}
\end{equation}
Coefficients $J_n$ are stored in column vector $\matr{J}_\mathrm{eq}$.

The tangential electric field on ${\cal S}$ in the equivalent configuration is enforced to be the same as in the original problem, $\nhat \times \vect{E}(\vect{r})$~\cite{DSA01, UTK_AWPL2017}.
The altered tangential magnetic field, $\nhat\times\vect{H}_\mathrm{eq}(\vect{r})$, is expanded with RWG basis functions, and its expansion coefficients are stored in vector $\matr{H}_{\mathrm{eq}}$.
In accordance with the equivalence principle, $\vect{J}_\mathrm{eq}(\vect{r})$ can be related to the original and modified tangential magnetic fields as
\begin{equation}
	\vect{J}_\mathrm{eq}(\vect{r}) = \nhat \times \left[ \vect{H}_\mathrm{eq}(\vect{r}) - \vect{H}(\vect{r}) \right]\,,
	\label{eq:equivalence1}
\end{equation}
whose discrete version reads
\begin{equation}
	\matr{J}_\mathrm{eq} = \matr{H}_\mathrm{eq} - \matr{H}\,.
	\label{eq:equivalence2}
\end{equation}

We again relate the tangential electric and magnetic fields on ${\cal S}$ in the equivalent problem by the Stratton-Chu formulation, this time invoking both the electric field integral equation (EFIE) and the MFIE. Discretizing as before and testing with rotated RWG functions, we get
\begin{align}
	j\omega\mu_l{\matr{L}}_{l} {\matr{H}_\mathrm{eq}} - {\matr{K}}_{l} \matr{E} &= 0, \label{eq:EFIEo}\\
	\eta_l{\matr{K}}_{l} {\matr{H}_\mathrm{eq}} + \eta_lj\omega\epsilon_l{\matr{L}}_{l} \matr{E} &= 0 \label{eq:MFIEo}.
\end{align}
Operators $\matr{L}_{l}$ and $\matr{K}_{l}$ are computed with the material properties of the $l^{\mathrm{th}}$ dielectric layer.
Since $\matr{L}_{l}$ is well-conditioned while $\matr{K}_{l}$ is not, the combined field integral equation (CFIE) is formed by adding equal parts of~\eqref{eq:EFIEo} and~\eqref{eq:MFIEo}, to ensure that both electric and magnetic fields are well-tested~\cite{Gibson2014}. This yields
\begin{align}
	{\matr{G}}_{J} {\matr{H}_\mathrm{eq}} + {\matr{G}}_{M} \matr{E} = 0\,,
	\label{eq:StrattonChu_Discrete2}
\end{align}
where
\begin{align}
	\matr{G}_{J} = 0.5\left(j\omega\mu_l\matr{L}_{l} + \eta_l\matr{K}_{l}\right) \label{eq:GJ}\\
	\matr{G}_{M} = 0.5\left(j\omega\epsilon_l\eta_l\matr{L}_{l} - \matr{K}_{l}\right). \label{eq:GM}
\end{align}
%
We explicitly relate the tangential magnetic field to the tangential electric field via the surface admittance operator $\matr{Y}_\mathrm{out}$ as
\begin{equation}
\matr{H}_\mathrm{eq} = \underbrace{-\matr{G}_{J}^{-1} \matr{G}_{M}}_{\matr{Y}_\mathrm{out}} \matr{E}\,.
\label{eq:Y0}
\end{equation}
%
%
%
%

Finally, by substituting~\eqref{eq:Y0} and~\eqref{eq:Yin} into~\eqref{eq:equivalence2} we obtain
\begin{equation}
	\matr{J}_\mathrm{eq} = \underbrace{\left[ \matr{Y}_\mathrm{out} - \matr{Y}_\mathrm{in} \right]}_{\matr{Y}_{\Delta}} \matr{E}\,,
\end{equation}
where $\matr{Y}_{\Delta}$ is the differential surface admittance (DSA) operator that accurately models electromagnetic fields inside the conductor~\cite{DSA01, UTK_AWPL2017, UTK_EPEPS2017}. This operator does not require a volumetric mesh and is applicable to conductors of arbitrary geometry.

%
For a multi-conductor system with $P$ conductors, the procedure described above is individually applied to each conductor, leading to a block diagonal matrix relating conductor-wise tangential electric fields and equivalent current densities,
\begin{equation}
\underbrace{\begin{bmatrix} \matr{J}^{(1)} \\ \vdots \\ \matr{J}^{(N_c)}\end{bmatrix}}_{\matr{J}_\mathrm{eq}} = \underbrace{\begin{bmatrix} \matr{Y}_{\Delta}^{(1)} & & \\ & \ddots & \\ & & \matr{Y}_{\Delta}^{(N_c)}  \end{bmatrix}}_{\matr{Y}_{\Delta}}
\underbrace{\begin{bmatrix} \matr{E}^{(1)} \\ \vdots \\ \matr{E}^{(N_c)} \end{bmatrix}}_{\matr{E}}\,. \label{JYsE}
\end{equation}

\subsection{Exterior Problem via A-EFIE} \label{sec:exterior}
We can now obtain an integral equation for tangential fields in the exterior problem to capture the coupling between different conductors.
%
%
A well-known pitfall of EFIE-based MoM formulations is low-frequency breakdown~\cite{lfbreakdown}.
Since this issue occurs due to the imbalance between scalar and vector potential at low frequencies, it can be mitigated by separating current and charge densities in the EFIE~\cite{aefie},
\begin{multline}
	\vect{E}(\vect{r}) + j\omega\mu_0\,\int_{S'}{\dyadic{G}(\vect{r},\vecprime{r})\cdot\vect{J}_\mathrm{eq}(\vecprime{r})\,dS'}\\ + \epsilon_0^{-1}\grad\int_{S'}{{G_\phi}(\vect{r},\vecprime{r})\,\rho_s(\vecprime{r})\,dS'} = \vect{E}_\mathrm{inc}, \label{MPIE}
\end{multline}
where $\rho_s(\vecprime{r})$ is the surface charge density on mesh triangles, and $\mu_0$ and $\epsilon_0$ are the permeability and permittivity of free space. The term $\dyadic{G}$ is the dyadic part of the MGF, and $G_\phi$ is its scalar part. Computation of the MGF is expounded in \secref{sec:MGF}.

Current and charge density can also be related via the continuity equation~\cite{aefie},
\begin{equation}
	\divg\vect{J}_\mathrm{eq}(\vecprime{r}) + j\omega\rho_s(\vecprime{r}) = 0. \label{continuityEM}
\end{equation}
Charge density is discretized using area-normalized pulse basis functions, $h_n(\vecprime{r})$,
\begin{align}
	{\rho}_s(\vecprime{r}) &= \sum_{n=1}^{N_t} \rho_n h_{n}(\vecprime{r}) \label{eq:rhoout}\,,
\end{align}
and coefficients $\rho_n$ are stored in vector $\matr{\rho}$. Term $N_t$ is the total number of mesh triangles. Equations~\eqref{MPIE} and~\eqref{continuityEM} can now be discretized to yield an augmented EFIE system~\cite{aefie}
\begin{equation}
\setlength\arraycolsep{3pt}
		\begin{bmatrix}
		\matr{Z}_{\mathrm{EM}} & -\matr{D}^T\matr{Z}_\Phi \\ 
		\matr{D} & jk_0\matr{I} \\ 
		\end{bmatrix}
		\begin{bmatrix}
		\matr{J}_\mathrm{eq} \\ c_0\matr{\rho}
		\end{bmatrix} = 
		\begin{bmatrix}
		\matr{E}_\mathrm{inc} \\ \matr{0}
		\end{bmatrix}, \label{systemEM}
\end{equation}
where $c_0$ and $k_0$ are the speed of light and wave number in vacuum, respectively, and
\begin{equation}
	\matr{Z}_{\mathrm{EM}} = jk_0\matr{L}_m + \dfrac{1}{\eta_{0}}\matr{I}_{\mathrm{orth}}\cdot\matr{Y}_\Delta^{-1}.  \label{V_J}
\end{equation}
Operator $\matr{L}_m$ is now built with the MGF using the proposed technique described in \secref{sec:MGF}.
The entries of matrix block $\matr{Z}_\Phi$ are provided in literature~\cite{aefie}. The only difference in this case is that the scalar part of the MGF is used as the kernel, using the proposed technique.
Matrix $\matr{I}$ is the identity matrix, and $\matr{D}$ is an incidence matrix that acts as a spatial derivative operator, as defined previously~\cite{chew1}. Matrix $\matr{I}_{\mathrm{orth}}$ represents the projection of rotated RWG basis functions on RWG functions.
%
Coupling of the augmented system \eqref{systemEM} to ports via Th\'evenin equivalent circuits is described in the Appendix. Enforcement of charge neutrality and preconditioning of the final system of equations are also provided in the Appendix.

\subsection{Conductors Traversing Layers} \label{sec:continuity}
Computing the DSA operator requires that a conductor be entirely situated in a single dielectric layer. To allow for this without sacrificing generality, we enforce continuity conditions that enable conductors to be in contact with each other within or across layers.

Continuity of currents is enforced by modifying the continuity equation for a pair of contact triangles as
\begin{equation}
	\divg\vect{J}_\mathrm{eq,1}(\vecprime{r}) + \divg\vect{J}_\mathrm{eq,2}(\vecprime{r}) + j\omega\rho_{s,12}(\vecprime{r}) = 0, \label{continuity_contact}
\end{equation}
where $\vect{J}_\mathrm{eq,1}(\vecprime{r})$ is the equivalent current density on the contact triangle of one of the conductors, and $\vect{J}_\mathrm{eq,2}(\vecprime{r})$ is the current density on its counterpart. The term $\rho_{s,12}(\vecprime{r})$ represents the total charge density shared by the pair of contact triangles. This equation manifests itself only as a slight modification to the matrix $\matr{D}$ in \eqref{systemEM}.

\section{Accelerated Computation of the MGF} \label{sec:MGF}
When accelerated with techniques such as AIM and FMM, interactions between source and field points are separately computed for points close to each other (``near-region'') and points far from each other (``far-region'')~\cite{AIMbles, FMA}.
One of the computational bottlenecks is near-region matrix fill, and the proposed method is developed to specifically address this phase.

\subsection{Near-Region MGF Computation} \label{sec:mgfform}
We consider expressions for the MGF proposed in formulation C of Michalski and Zheng \cite{MGF02}, which consists of a dyadic term $\dyadic{G}$ and a scalar term $G_\phi$. However, it should be noted that the method presented here is extensible to any valid formulation.
Assuming that the dielectric layers are stacked along the $z$ axis, the dyadic term is
\begin{equation}
    \dyadic{G} = 
    \begin{bmatrix}
    	G_{xx} & 0 & G_{xz} \\
		0 & G_{yy} & G_{yz} \\
		G_{zx} & G_{zy} & G_{zz} \\
    \end{bmatrix}. \label{dyadicG}
\end{equation}
Each component of $\dyadic{G}$, as well as $G_\phi$, has the general form~\cite{MGF02}
\begin{multline}
	G\left(k, \vect{r}, \vect{r}\,'\right) =\\
	C_l\left(\rho, \epsilon_l, \mu_l\right)\int_0^\infty dk_\rho J_\nu\left(\rho k_\rho\right)\widetilde{G}\left(k_\rho, z, z'\right)k^{\nu+1}_\rho. \label{MGF}
\end{multline}
Function $J_\nu\left(\rho k_\rho\right)$ is the Bessel function of first kind and order $\nu$, and $\widetilde{G}\left(k_\rho, z, z'\right)$ is the spectral MGF corresponding to $G\left(k, \vect{r}, \vect{r}\,'\right)$. Quantity $k$ is the wave number, $k_\rho$ is the wave number in the lateral ($xy$) plane,
and $\rho = \sqrt{(x - x')^2 + (y - y')^2}$. The Bessel function of order $\nu = 0$ is used for diagonal components and for $G_\phi$, while $\nu = 1$ is used for off-diagonal components. Constant $C_l$ depends on $\rho$ and on the (complex) permittivity $\epsilon_l$ and permeability $\mu_l$ of layer $l$. In what follows, we suppress $C_l$ for brevity, since its presence does not change the derivation of the proposed method. Note that the path of integration can be taken along the real $k_\rho$ axis, except a detour at small $k_\rho$ values to avoid surface wave poles~\cite{SI_PE}. Except in the detour, $k_\rho$ is a real number.

In order to speed up convergence of Sommerfeld integrands, the quasistatic (QS) contribution, $\widetilde{G}_\mathrm{QS}\left(k_\rho, z, z'\right)$, can be extracted from spectral domain functions $\widetilde{G}\left(k_\rho, z, z'\right)$.
Since analytical expressions for QS terms are available~\cite{qse}, their contribution is added back analytically in the spatial domain. Thus we can define
\begin{subequations}
	\begin{align}
	\widetilde{G}_\mathrm{E}\left(k_\rho, z, z'\right) &\triangleq \widetilde{G}\left(k_\rho, z, z'\right) - \widetilde{G}_\mathrm{QS}\left(k_\rho, z, z'\right), \\
	G_\mathrm{E}\left(k, \vect{r}, \vect{r}\,'\right) &\triangleq G\left(k, \vect{r}, \vect{r}\,'\right) - G_\mathrm{QS}\left(k, \vect{r}, \vect{r}\,'\right),
	\end{align}
\end{subequations}
where subscript $\mathrm{QS}$ denotes quasistatic contributions, and subscript $\mathrm{E}$ indicates remainders after extracting QS terms. Extraction of QS terms leads to faster decay of the integrand in~\eqref{MGF}. This allows us to truncate the semi-infinite integral at some finite $k_{\rho0}$, beyond which $\widetilde{G}_\mathrm{E}\left(k_\rho, z, z'\right)$ is negligible in comparison to $\widetilde{G}_\mathrm{QS}\left(k_\rho, z, z'\right)$. Selection of $k_{\rho0}$ is discussed in \secref{sec:truncation}. This allows us to write 
\begin{equation}
	G_\mathrm{E}\left(k, \vect{r}, \vect{r}\,'\right) = \int_0^{k_{\rho0}} dk_\rho J_\nu\left(\rho k_\rho\right)\widetilde{G}_\mathrm{E}\left(k_\rho, z, z'\right)k^{\nu+1}_\rho. \label{MGFtrunc}
\end{equation}

Since the proposed technique is applied to near-region computations only, and $k_\rho$ is bounded above by $k_{\rho0}$, the argument of $J_\nu\left(\rho k_\rho\right)$ in~\eqref{MGFtrunc} is small.
The Bessel function can then be expanded as a Taylor series centered at $\rho k_\rho  = 0$~\cite{abrstegun},
\begin{equation}
	J_\nu\left(\rho k_\rho\right) = \left(0.5\rho k_\rho\right)^\nu\sum_{i = 0}^{\infty}{\dfrac{\left(-0.25\rho^2k_\rho^2\right)^i}{i!\left(\nu + i\right)!}}. \label{Jasc}
\end{equation}
This expansion allows us to write the Bessel function as a product of terms that depend only on $\rho$, and terms that depend only on frequency via $k_\rho$,
\begin{equation}
	J_\nu\left(\rho k_\rho\right) = \sum_{i = 0}^{\infty}{\rho^{\nu+2i}k_\rho^{\nu+2i}\dfrac{\left(0.5\right)^{\nu}\left(-0.25\right)^i}{i!\left(\nu + i\right)!}}. \label{Jasc2}
\end{equation}
Inserting the expansion into~\eqref{MGFtrunc} yields
\begin{multline}
	G_\mathrm{E}\left(k, \vect{r}, \vect{r}\,'\right) =\\
	 \int_0^{k_{\rho0}} dk_\rho\, \widetilde{G}_\mathrm{E}\left(k_\rho, z, z'\right) k^{\nu+1}_\rho \\
	 \sum_{i = 0}^{\infty} \rho^{\nu+2i}k_\rho^{\nu+2i}\dfrac{\left(0.5\right)^{\nu}\left(-0.25\right)^i}{i!\left(\nu + i\right)!}. \label{MGFseries}
\end{multline}
Since the argument of the Bessel function is expected to be small, only a few terms are required in its expansion, as discussed in \secref{sec:truncation}.
%
%
This allows us to truncate the summation in~\eqref{MGFseries} at a finite number of terms $N_J$,
\begin{multline}
	G_\mathrm{E}\left(k, \vect{r}, \vect{r}\,'\right) =\\
	\int_0^{k_{\rho0}} dk_\rho\, \widetilde{G}_\mathrm{E}\left(k_\rho, z, z'\right) k^{\nu+1}_\rho \\
	\sum_{i = 0}^{N_J} \rho^{\nu+2i}k_\rho^{\nu+2i}\dfrac{\left(0.5\right)^{\nu}\left(-0.25\right)^i}{i!\left(\nu + i\right)!}. \label{MGFseries3}
\end{multline}
%
Numerical verification of the validity of this truncation is provided in \secref{Jnum}.

The main advantage of expressing the Sommerfeld integrals in this way is that terms dependent on $\rho$ are separated from terms that depend on frequency. Since both the summation and integration are now finite, their order can be swapped. This allows $\rho$-dependent terms to be extracted from the integral to get
\begin{multline}
	G_\mathrm{E}\left(k, \vect{r}, \vect{r}\,'\right) =\\
	\sum_{i = 0}^{N_J} \dfrac{\left(0.5\right)^{\nu}\left(-0.25\right)^i}{i!\left(\nu + i\right)!} \rho^{\nu+2i} \int_0^{k_{\rho0}} dk_\rho\, \widetilde{G}_\mathrm{E}\left(k_\rho, z, z'\right) k_\rho^{2\nu+2i+1}. \label{MGFseries4}
\end{multline}

The integrand in~\eqref{MGFseries4} now only depends on the simulation frequency and coordinates along the direction of stratification. The integral can thus be precomputed for a predetermined set of $z$--$z'$ pairs rather than every possible combination of source and observation points. This provides significant savings in the computation time of MoM matrix entries. Precomputation of the integrals is particularly advantageous for on-chip structures, which are often much smaller along the direction of stratification than in the lateral directions.

Once the integrals are precomputed, one can use any suitable interpolation technique to approximate the integral at any $z$--$z'$ pair. In this work, we precompute the integrals for a uniform set of $z$--$z'$ pairs and simply use the value at the nearest neighbour during the MoM matrix fill.
Precomputation points are chosen such that there are approximately three $z$--$z'$ pairs per vertically-oriented mesh triangle.

The proposed method provides three significant advantages over DCIM:
\begin{enumerate}
	\item In DCIM, the number of images required varies with different layer configurations and frequencies. Complicated stack-ups may require in excess of $10$--$15$ images, each of which involves computing a complex exponential factor. The matrix assembly cost is proportional to this number. With the proposed series expansion, it is demonstrated in \secref{sec:results} that $N_J \sim 3$ is sufficient to obtain a good approximation for on-chip applications.
	\item As a result of the integral precomputation, only simple algebraic operations are required in~\eqref{MGFseries4} during MoM matrix fill. This is an important computational advantage, since it eliminates the need to compute relatively expensive exponentials when calculating source-observation interactions.
	\item The series expansion allows for direct error control by picking an appropriate number of expansion terms, as discussed in \secref{sec:truncation}. In DCIM, errors are controlled indirectly and heuristically during the fitting procedure, unless a more rigorous approach, such as the one suggested in~\cite{DCIM02}, is applied.
\end{enumerate}
%

\subsection{Truncation Point Selection} \label{sec:truncation}

An important consideration is the selection of $k_{\rho0}$ and $N_J$ to obtain an accurate approximation of $G_\mathrm{E}\left(k, \vect{r}, \vect{r}\,'\right)$.
The integral truncation point $k_{\rho0}$ is obtained by picking a tolerance $\delta_{k_{\rho0}}$ such that 
\begin{equation}
	\dfrac{\abs{\widetilde{G}_\mathrm{E}\left(k_\rho, z, z'\right)}}{\abs{\widetilde{G}_\mathrm{QS}\left(k_\rho, z, z'\right)}} < \delta_{k_{\rho0}} \quad \forall\,k_\rho > k_{\rho0}.
\end{equation}
The resulting $k_{\rho0}$ represents the value of $k_\rho$ beyond which $\widetilde{G}_\mathrm{E}\left(k_\rho, z, z'\right)$ can be neglected in comparison to the contribution of QS terms, which are added analytically in spatial domain. In this paper, we use $\delta_{k_{\rho0}} = 10^{-2}$ in all test cases.

The number of expansion terms can now be chosen by picking a tolerance $\delta_{N_J}$ such that
\begin{equation}
	\dfrac{\abs{J_\nu\left(\rho k_{\rho}\right) - \widetilde{J}_{\nu, N_J}\left(\rho k_{\rho}\right)}}{\max\limits_{\rho k_{\rho}} \abs{J_\nu\left(\rho k_{\rho}\right)}} < \delta_{N_J} \quad \forall\, \rho k_\rho \in [0, \rho_{0} k_{\rho0}],
\end{equation}
where $\rho_{0}$ is the diameter of the near-region, and $\widetilde{J}_{\nu, N_J}\left(\rho k_{\rho}\right)$ is the Taylor approximation of $J_\nu\left(\rho k_{\rho}\right)$ using $N_J$ terms,
\begin{equation}
	\widetilde{J}_{\nu, N_J}\left(\rho k_{\rho}\right) = \sum_{i = 0}^{N_J-1}{\rho^{\nu+2i}k_\rho^{\nu+2i}\dfrac{\left(0.5\right)^{\nu}\left(-0.25\right)^i}{i!\left(\nu + i\right)!}}. \label{Jasc3}
\end{equation}
In this paper, we use $\delta_{N_J} = 10^{-2}$ in all test cases.
This procedure may yield different values of $N_J$ for different $z$--$z'$ pairs, and different MGF components. We use the largest value of $N_J$ computed. In all test cases considered here, the maximum value of $N_J$ required was $3$.

\subsection{Numerical Analysis}\label{Jnum}

\begin{table}[t]
	\centering
	\caption{Dielectric layer configuration for numerical verification~in~\secref{Jnum}~and~the interconnect~network~in~\secref{sec:results:interconnect}.}
	\begin{tabular}{llll}
		\toprule
		$\epsilon_r$ & $\mu_r$ & $\sigma$ (S/m) & Height ($\mu$m) \\
		\midrule
		11.5 & 1.0 & 0.01 & 3 \\
		9.8 & 1.0 & 0.001 & 4 \\
		12.5 & 1.0 & 0.1 & 10 \\
		6.0 & 1.0 & 0.0001 & 9 \\
		4.4 & 1.0 & 0.0 & 4 \\
		\bottomrule
	\end{tabular}
	\label{Jnumcases}
\end{table}

\begin{figure}[t]
	\subfloat[][]
	{
		\includegraphics[width=9cm, page=1, trim={0cm 0.0cm 0cm 0.5cm}, clip]{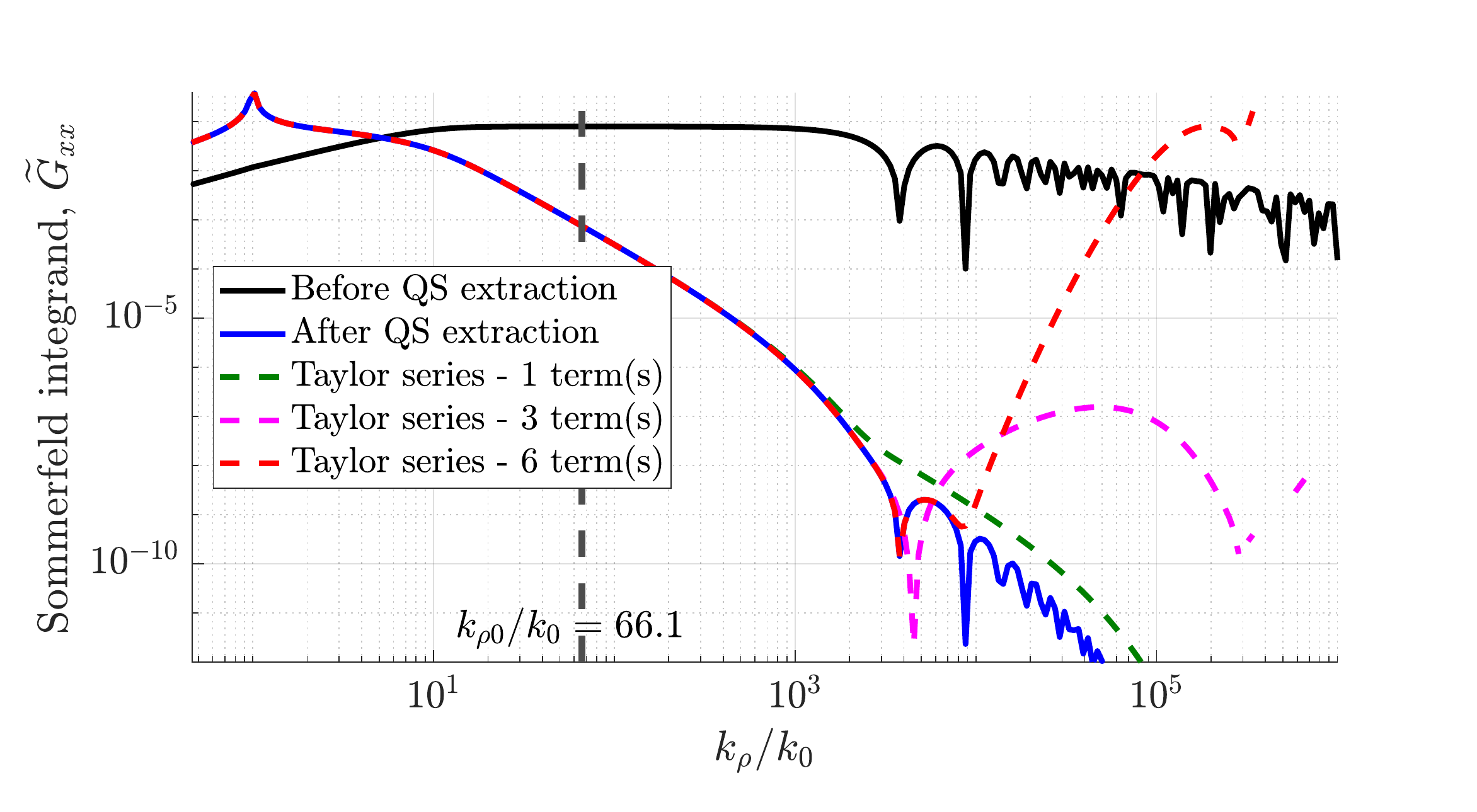}
		\label{Jcase1a}
	}
	\\
	\subfloat[][]
	{
		\includegraphics[width=9cm, page=1, trim={0cm 0.0cm 0cm 0.9cm}, clip]{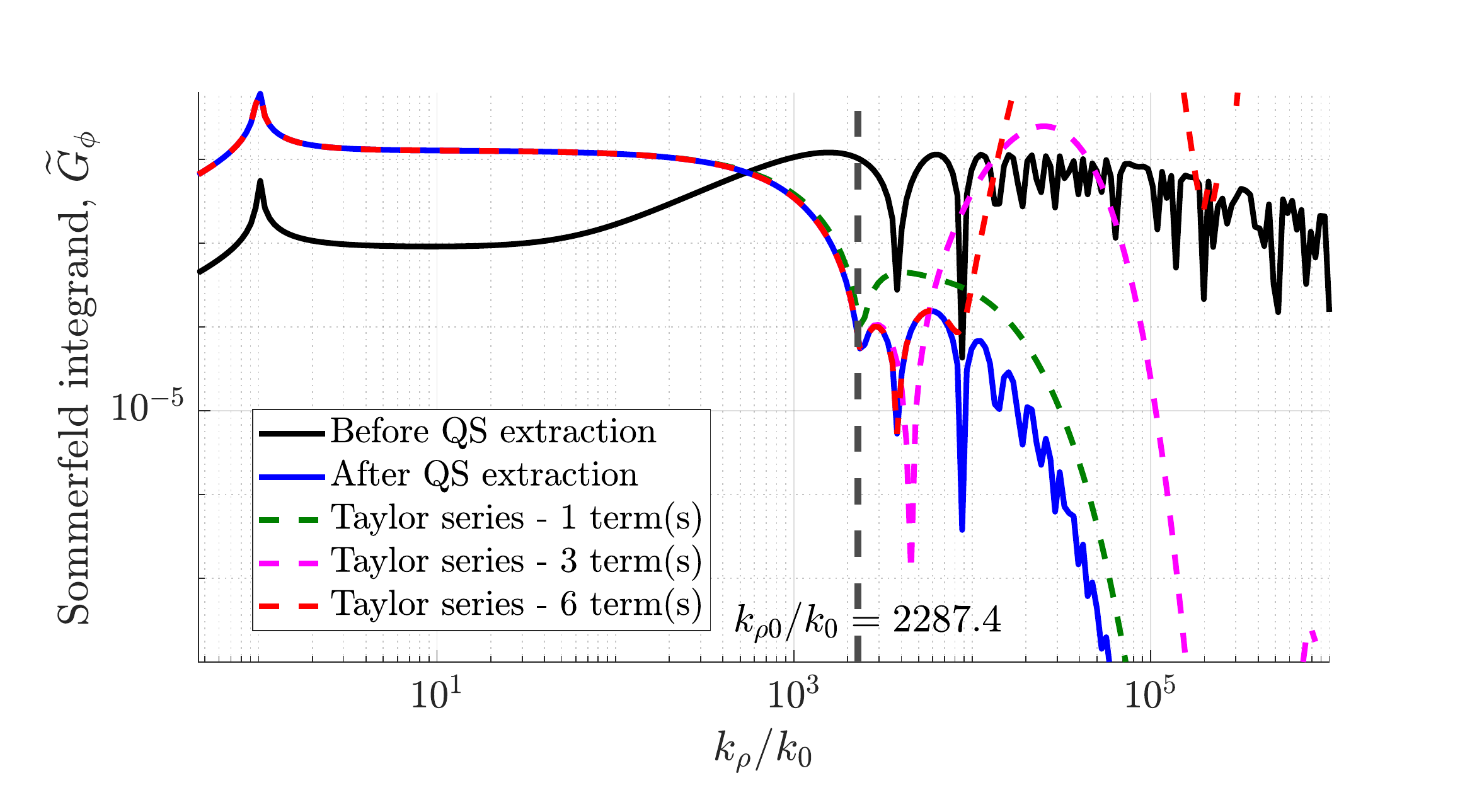}
		\label{Jcase1b}
	}
	\\
	\subfloat[][]
	{
		\includegraphics[width=9cm, page=1, trim={0cm 0.0cm 0cm 0.9cm}, clip]{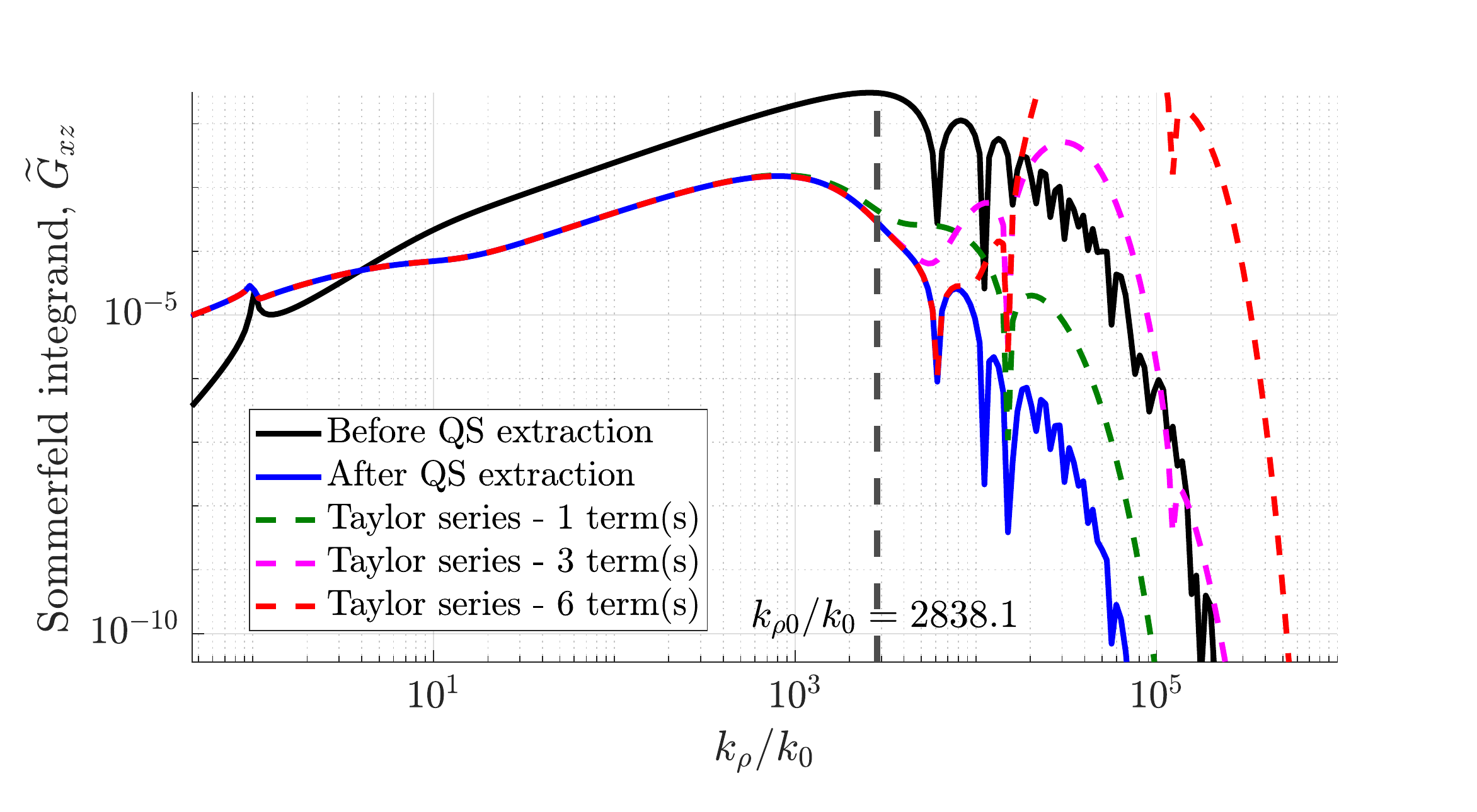}
		\label{Jcase1c}
	}
	\caption{Behavior of Sommerfeld integrands before and after QS extraction, and after series approximation: (a) Component $\widetilde{G}_{xx}$, $z = z' = 21\,\mu$m, (b) Component $\widetilde{G}_{\phi}$, $z = z' = 21\,\mu$m, and (c) Component $\widetilde{G}_{xz}$, $z = 21\,\mu$m, $z' = 17\,\mu$m. Each plot corresponds to the stack-up in \tabref{Jnumcases}.}
	\label{Jcases1}
\end{figure}

To validate the proposed method, we analyze the behaviour of the Sommerfeld integrand for three cases:
\begin{itemize}
	\item integrand in~\eqref{MGF} for the original kernel $\widetilde{G}\left(k_\rho, z, z'\right)$,
	\item integrand in~\eqref{MGF} with kernel, $\widetilde{G}_\mathrm{QS}\left(k_\rho, z, z'\right)$, and
	\item integrand in~\eqref{MGFseries3} for different $N_J$.
\end{itemize}
The analysis is performed on the stack-up described in \tabref{Jnumcases}, which is used for the realistic interconnect network considered in \secref{sec:results:interconnect}. A near-region diameter of $\rho_\mathrm{0} = 30\,\mu$m is assumed, which is the actual value used for simulating the aforementioned interconnect network. Three representative cases for different MGF components are shown in \figref{Jcases1}, each computed at $1\,$GHz. Vertical dashed lines indicate the integral truncation point $k_{\rho0}$, as computed via the procedure in \secref{sec:truncation}.

As expected, the plots confirm that extraction of QS terms causes the Sommerfeld integrand to decay significantly faster, making them suitable for application of the proposed series expansion for relatively small $N_J$. The number of expansion terms required was $1$, $3$ and $3$ for the components in \figref{Jcase1a}, \figref{Jcase1b} and \figref{Jcase1c}, respectively.

The accuracy of the proposed method is further confirmed by analyzing the values and relative errors in computing the spatial domain MGF, $G\left(k, \vect{r}, \vect{r}\,'\right)$.
\tabref{Intvals} compares the reference result of full numerical integration as per~\eqref{MGF}, with the result of the proposed method. Each of the three test cases corresponds to the spectral domain plots in \figref{Jcases1}. It is clear that the proposed method yields accurate values with a relative error below $0.4\%$ in all test cases, which decreases for increasing $N_J$.

\begin{table*}[t]
	\centering
	\caption{Values of Sommerfeld integrals for~validation~in~\secref{Jnum}.}
	\begin{tabular}{lllllll}
		\toprule
		& \multicolumn{2}{c}{${G}_{xx}$, \figref{Jcase1a}} & \multicolumn{2}{c}{${G}_{\phi}$, \figref{Jcase1b}} & \multicolumn{2}{c}{${G}_{xz}$, \figref{Jcase1c}} \\
		\cmidrule{2-3} 
		\cmidrule{4-5}
		\cmidrule{6-7}
		& Integral value & Rel. error ($\%$) & Integral value & Rel. error ($\%$) & Integral value & Rel. error ($\%$)\\
		\midrule
		Equation~\eqref{MGF} & $9.900 - 484.301j$ & -- & $-20.439 + 1547.192j$ & -- & $6.275 - 1846.273j$ & -- \\
		Proposed, $1$ term & $9.198 - 484.321j$ & $0.145$ & $-22.511 + 1541.879j$ &  $0.369$ & $6.926 - 1841.268j$  & $0.273$ \\
		Proposed, $2$ terms & $9.197 - 484.322j$ & $0.145$ & $-20.079 + 1552.900j$ & $0.370$ & $5.966 - 1849.434j$  & $0.172$ \\
		Proposed, $3$ terms & $9.197 - 484.322j$ & $0.145$ & $-20.258 + 1552.471j$ & $0.341$ & $6.051 - 1848.715j$ & $0.133$\\
		\bottomrule
	\end{tabular}
	\label{Intvals}
\end{table*}

To ensure that the proposed method yields accurate spatial domain MGF values in the entire near-region, we compare the spatial domain results of the proposed method with full numerical integration, and the contribution of QS terms. Two representative test cases are described below.

%

\begin{table}[t]
	\centering
	\caption{Dielectric layer configuration for numerical verification~in~\secref{Jnum}~and~the inductor~coil~in~\secref{sec:results:inductor}.}
	\begin{tabular}{lllc}
		\toprule
		$\epsilon_r$ & $\mu_r$ & $\sigma$ (S/m) & Height ($\mu$m) \\
		\midrule
		2.1 & 1.0 & 0.0 & 50 \\
		12.5 & 1.0 & 0.0 & 50 \\
		PEC & -- & -- & -- \\
		\bottomrule
	\end{tabular}
	\label{Jnumcases2}
\end{table}

\begin{figure}[t]
	\subfloat[][]
	{
		\includegraphics[width=9cm, page=1, trim={0cm 0cm 0cm 0cm}, clip=false]{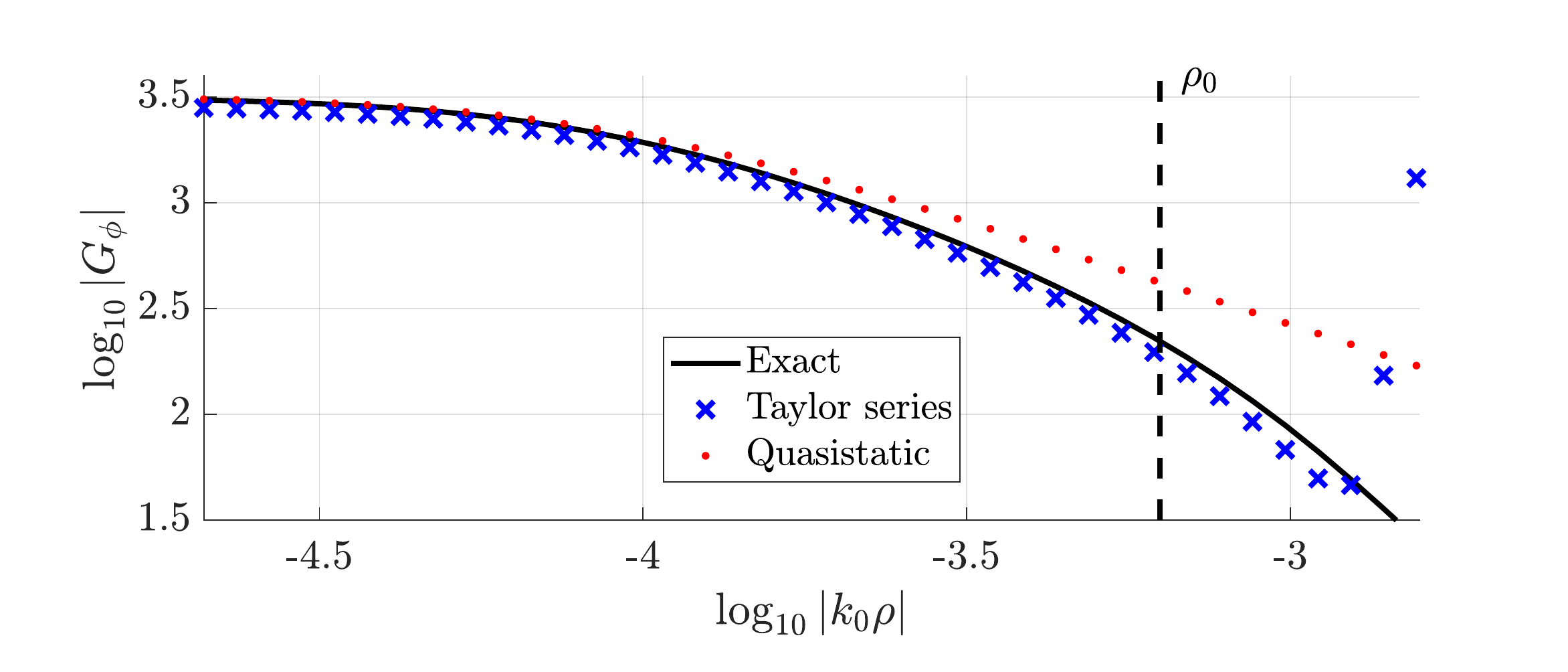}
		\label{Gcase1}
	}
	\\
	\subfloat[][]
	{
		\includegraphics[width=9cm, page=1, trim={0cm 0cm 0cm 0cm}, clip=false]{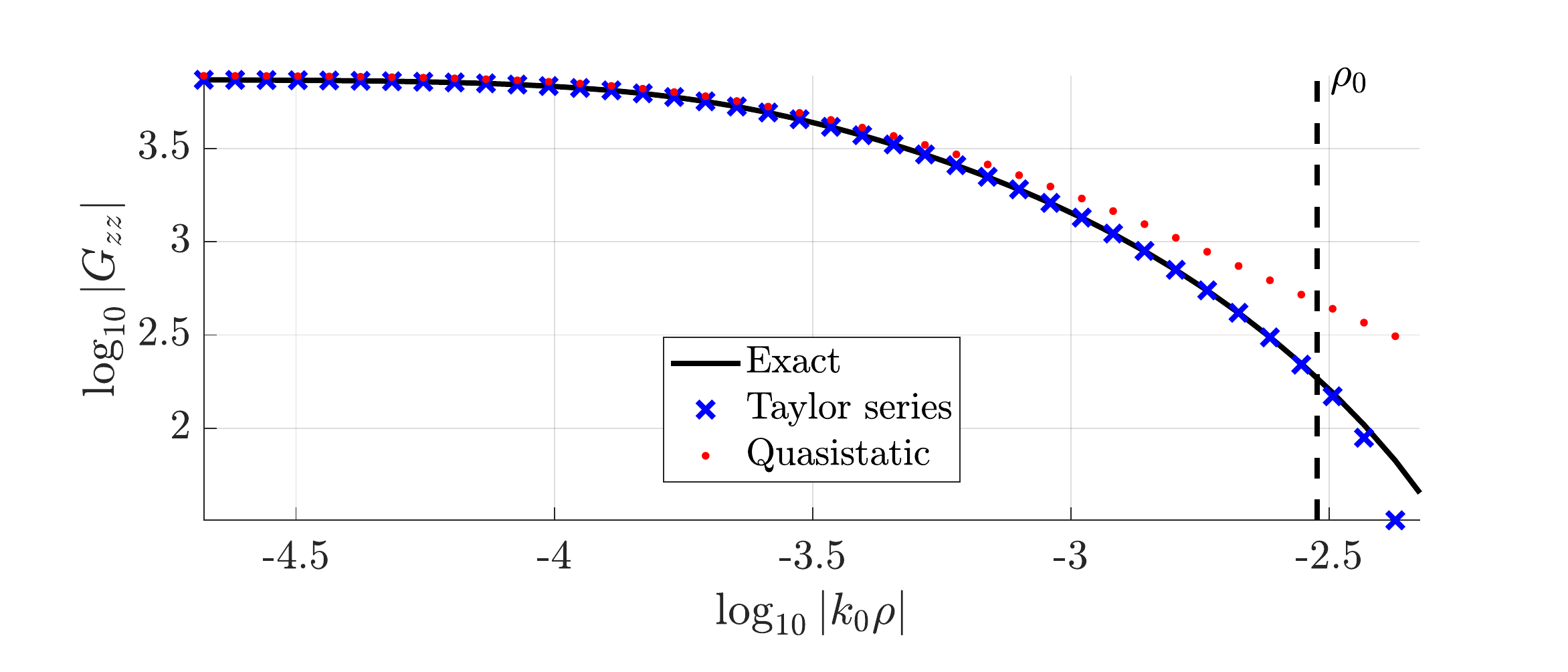}
		\label{Gcase2}
	}
	\caption{Validation of spatial domain MGF components, with quasistatic contributions, for (a) $G_{\phi}$ with the stack-up in~\tabref{Jnumcases}, $z = 17\,\mu$m, $z' = 21\,\mu$m and (b) $G_{zz}$ with the stack-up in~\tabref{Jnumcases2}, $z = 29\,\mu$m, $z' = 18\,\mu$m.}
	\label{Gcases}
\end{figure}

\subsubsection{Component $G_{\phi}$, stack-up in \tabref{Jnumcases}}

The scalar component of the MGF at $1\,$GHz, along with contributions of quasistatic terms and the series approximation, is plotted in \figref{Gcase1} for the stack-up in \tabref{Jnumcases}. In this case, $N_J = 3$ expansion terms were sufficient. The vertical dashed line indicates the near-region diameter ($\rho_0 = 30\,\mu$m) relevant to the interconnect network discussed in \secref{sec:results:interconnect}. Clearly, the series expansion is sufficient for use in the near-region, while the QS contribution deviates from correct MGF values well within the near-region.

\subsubsection{Component $G_{zz}$, stack-up in \tabref{Jnumcases2}}

The $zz$ component of the MGF at $1\,$GHz for the stack-up in \tabref{Jnumcases2} is validated in \figref{Gcase2}, again with $N_J = 3$. This stack-up corresponds to the inductor coil simulated in \secref{sec:results:inductor}. The near-region diameter in this case is $\rho_0 = 143\,\mu$m. Again, it is clear that the proposed method is accurate in the near-region, while the QS contribution is insufficient on its own.

\section{Acceleration with AIM} \label{sec:AIM}

\begin{figure}[t]
	\subfloat[][] {
		\includegraphics[width=9cm, page=1, trim={0 0 0 0}, clip]{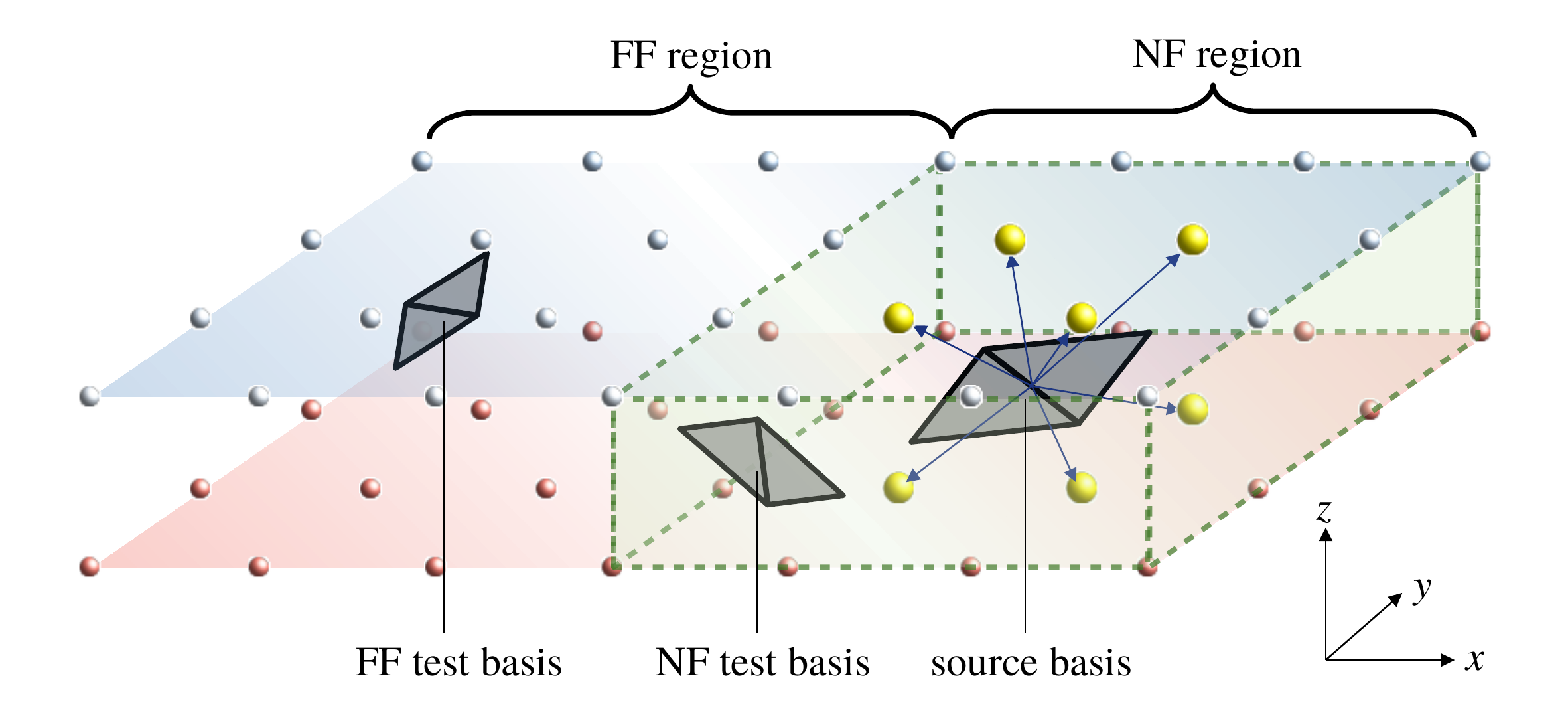} \label{AIM_grida}} \\
	\subfloat[][] {
		\includegraphics[width=9cm, page=2, trim={0 2cm 0 2.2cm}, clip]{figures/AIM_grid/AIM_grid_v3.pdf} \label{AIM_gridb} } \\
	\subfloat[][] {
		\includegraphics[width=9cm, page=3, trim={0 1.8cm 0 2.2cm}, clip]{figures/AIM_grid/AIM_grid_v3.pdf}  \label{AIM_gridc} }
	\caption{AIM procedure. (a) Current and charge density unknowns are projected from mesh to grid points. The box bounded by dashed lines represents the near-region (NF) of the source basis function, while the remaining volume corresponds to the far-region (FF). (b) Interactions between source and test grid points are accelerated with 2D FFTs to compute grid potentials. (c) Grid potentials are interpolated from grid points back on to the original mesh.}
	\label{AIM_grid}
\end{figure}

Assembling MoM matrices for large problems with hundreds of thousands of unknowns would require prohibitively large amounts of memory and CPU time. Moreover, factorizing large dense matrices using direct methods such as LU decomposition would be impractical for the same reasons.
We employ the adaptive integral method (AIM)~\cite{AIMbles}, modified to efficiently handle the MGF.
The principle behind AIM is to split the system matrix into near- and far-region interactions. Matrix elements corresponding to weakly-interacting basis functions in the far-region are projected onto a regular 3D grid, as shown in \figref{AIM_grida}. Fast Fourier transforms (FFTs) are leveraged to speed up the matrix-vector product in solving the final system of equations iteratively~\cite{pfftmain}. Grid point interactions are then interpolated back onto mesh basis functions, visualized in \figref{AIM_gridc}.

In the case of homogeneous media, the translation-invariance of the Green's function enables the use of 3D FFTs to accelerate interactions in all directions~\cite{pfftmain}.
However, the MGF is only translation-invariant in the lateral directions, and thus amenable to 2D FFTs along the $x$ and $y$ directions.
A 2D FFT-based method was initially proposed for the case of planar conductors lying in the $xy$ plane~\cite{AIM_MGF_2D}.
A technique combining 2D and 3D FFTs has been explored, but is only applicable when all source and observation points lie in one layer~\cite{convcorr_AIM}.
A more general procedure has also been proposed~\cite{okh_AIM_MGF_3D}, where 3D objects are modeled by projecting basis functions onto 2D stencils.
However, this requires a grid that is very dense along the direction of stratification ($z$) and makes the method more computationally expensive than the homogeneous version of AIM.

We propose an AIM-based procedure where basis functions are projected onto a 3D stencil, but grid point interactions are accelerated with 2D FFTs. This is unlike the 2D stencil-based approach~\cite{okh_AIM_MGF_3D}, and allows us to use a 3D grid whose spacing is the same as it would be in the homogeneous case.

In the proposed method, the projection and interpolation steps are similar to the homogeneous case~\cite{pfftmain}.
Interactions between grid points are computed by means of a convolution matrix, which encodes the Green's function at each grid point, with respect to a single source point. The main distinction between the proposed technique (which incorporates the MGF) and the homogeneous case is the structure and usage of the convolution matrix.
First, we define the following terms for convenience with reference to \figref{AIM_gridb}:
\begin{itemize}
	\item Each source grid point is represented by $\vect{r}_{i,j}^{\,m}$.
	\item Subscript $i = 1, 2 \ldots N_x$ are indices through nodes along the $x$ axis.
	\item Subscript $j = 1, 2 \ldots N_y$ are indices through nodes along the $y$ axis.
	\item Superscript $m = 1, 2 \ldots N_z$ is an index through each $xy$ source plane, along the $z$ axis.
	\item $N_x$, $N_y$ and $N_z$ are the total number of grid points in the $x$, $y$ and $z$ directions, respectively.
\end{itemize}

The translation invariance of the MGF along $x$ and $y$ means that one only needs to consider one source grid point per $xy$ plane. We can take $\{\vect{r}_{1,1}^{\,m}\}, m = 1, 2 \ldots N_z$ as the set of source grid points for each $xy$ grid plane. Each of these $N_z$ source grid points interacts with every grid point. Let $\vect{r}_{i,j}^{\,n}$ represent each test grid point, where $n = 1, 2 \ldots N_z$ is an index through each test layer. The total number of test grid points is $N_xN_yN_z$. Thus the total number of interactions to be computed is $N_z(N_xN_yN_z)$. This can more conveniently be understood as computing $N_z$ stacks of $N_x \times N_y \times N_z$ matrices, as shown in \figref{AIM_gridb}.

The entries of each matrix are the values of the MGF for that combination of source and test grid points. Since there are $8$ components in the MGF, including the scalar potential term, a total of $8$ such matrix stacks are to be computed. Each entry of the convolution matrix is given by $G_q(\vect{r}_{1,1}^{\,m}, \vect{r}_{i,j}^{\,n})$ for each $i$, $j$, $m$ and $n$, where $q = 1, 2 \ldots 8$ is an index through each MGF component. Once computed and arranged in two-level Toeplitz format, we can leverage 2D FFTs on each of the $N_z^2$ 2D grids to compute all possible grid interactions in the spatial frequency domain. The convolution matrix then has size $8 \times N_z \times N_x \times N_y \times N_z$.

This form of the convolution matrix requires $8N_z$ times more memory than the homogeneous counterpart. However, typical structures of interest, particularly in interconnect modeling, are significantly larger along the $x$ and $y$ directions than along $z$. In addition, the number of $z$ grid points needed here does not need to be any larger than in the homogeneous case. Computing several 2D FFTs is thus comparable, if not faster, than a single 3D FFT. Computing MGF entries in the convolution matrix is more expensive than the homogeneous counterpart, but this phase of the procedure is generally not among the most expensive of a MoM-based code. Thus the added memory and time costs are negligible in the context of an entire simulation, and inclusion of stratified media does not reduce the performance gains of AIM. This is quantitatively demonstrated in \secref{sec:results:interconnect}. Unlike previous works~\cite{AIM_MGF_2D, convcorr_AIM}, the proposed approach is valid for general 3D conductors embedded in multiple layers.

%

\section{Results} \label{sec:results}

The proposed technique is validated through two test cases, each of which was conducted on a $3.6$\,GHz desktop computer on a single thread.
\begin{table*}[t]
	\centering
	\caption{Performance statistics for each test case averaged over all frequency points.}
	\begin{tabular}{@{\extracolsep{4pt}}lllllllll}
		\toprule
		& \multicolumn{3}{c}{Inductor coil} & \multicolumn{3}{c}{Interconnect network}\\
		& \multicolumn{3}{c}{(\secref{sec:results:inductor})} &  \multicolumn{3}{c}{(\secref{sec:results:interconnect})} \\
		\cmidrule{2-4} 
		\cmidrule{5-7} 
		& Proposed & DCIM & HFSS & Proposed & DCIM & HFSS\\
		\midrule
		Mesh elements & $1,058$ & $1,058$ & $183,000$ & $32,876$ & $32,876$ & $872,516$ \\
		Memory used (GB) & $0.23$ & $0.23$ & $10.7$ & $3.3$ & $3.3$ & $71.4$ \\
		Near-region CPU time (s) & $10$ & $22$ & -- & $899$ & $1,934$ & -- \\
		Total CPU time (s) & $55$ & $66$ & $100$ & $2,656$ & $3,612$ & $3,233$ \\
		Average number of terms & $3$ & $7$ & -- & $3$ & $11$ & -- \\
		\bottomrule
	\end{tabular}
	\label{stats}
\end{table*}

\subsection{Inductor Coil} \label{sec:results:inductor}

\begin{figure}[t]
	\includegraphics[width=9cm, page=1, trim={0 0 0 13cm}, clip]{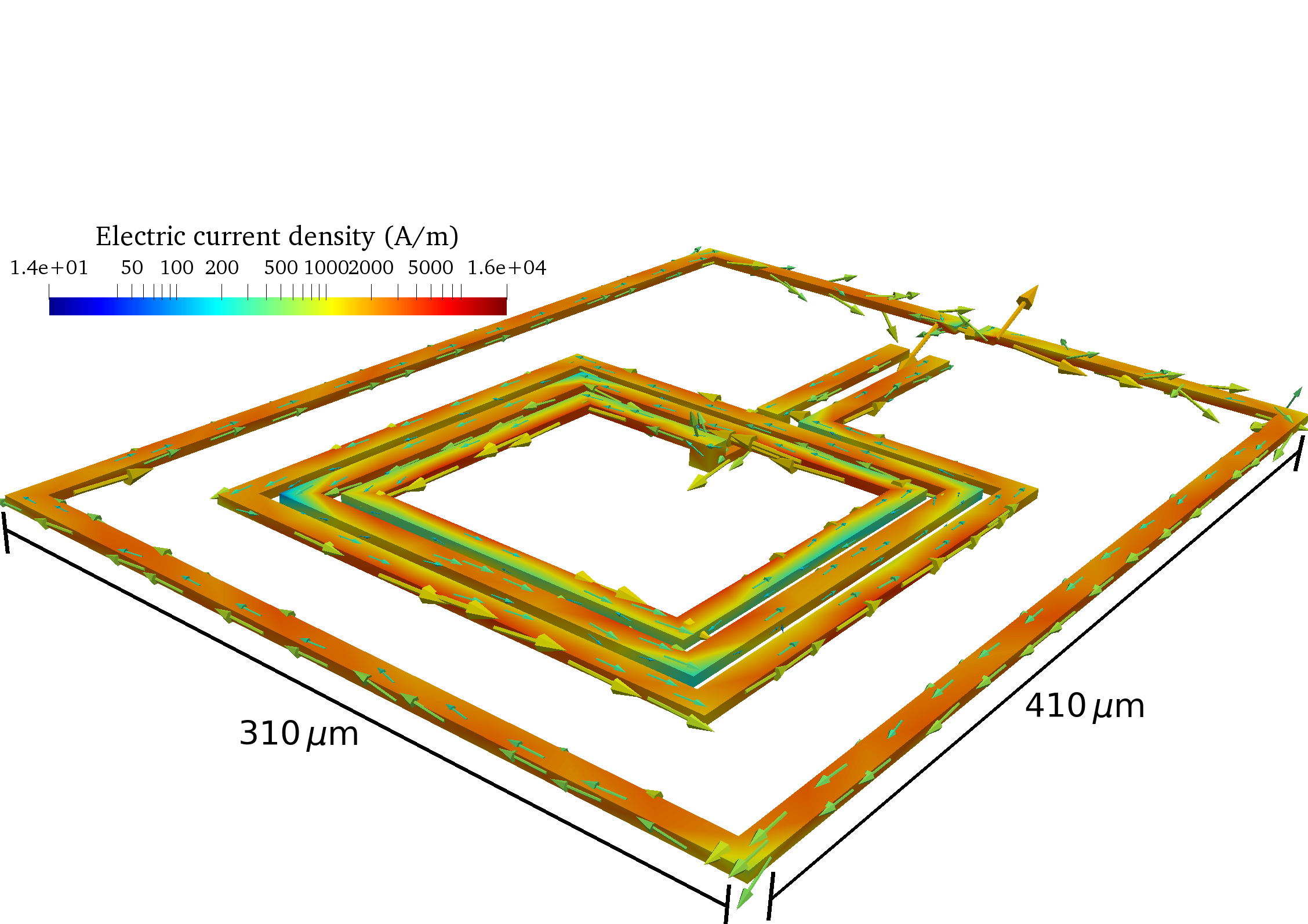}
	\caption{Computed surface current density distribution at $1$\,GHz on the two-port inductor coil in~\secref{sec:results:inductor}.}
	\label{ind_mult_v2_Js}
\end{figure}

First, we consider a two-port inductor~\cite{MITind} embedded in a stack of two dielectric layers whose configuration and properties are given in \tabref{Jnumcases2}. The layers are bounded by free space above, and by an infinite ground plane below. The trace cross section is $10\,\mu$m$ \times 4\,\mu$m. The inductor geometry and computed surface current density distribution at $1$\,GHz are shown in \figref{ind_mult_v2_Js}. Three terms were used in the Bessel function expansion for the MGF. Over the entire frequency range, on average $42$ iterations were required to reduce the relative residual norm below $10^{-6}$, for both DCIM and the proposed technique.

\begin{figure}[t]
	\subfloat[][]
	{
		\includegraphics[width=9cm, page=1, trim={0 0 0 0}, clip]{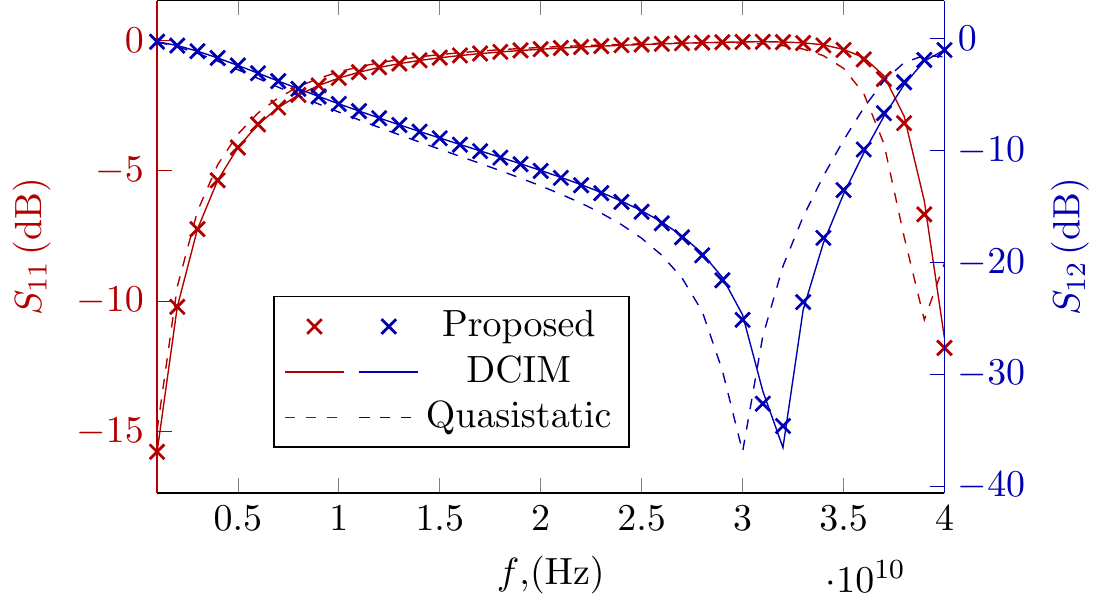}
		\label{inductor_multilayer_Slin}
	}\\
	\subfloat[][]
	{
		\includegraphics[width=9cm, page=1, trim={0 0 0 0}, clip]{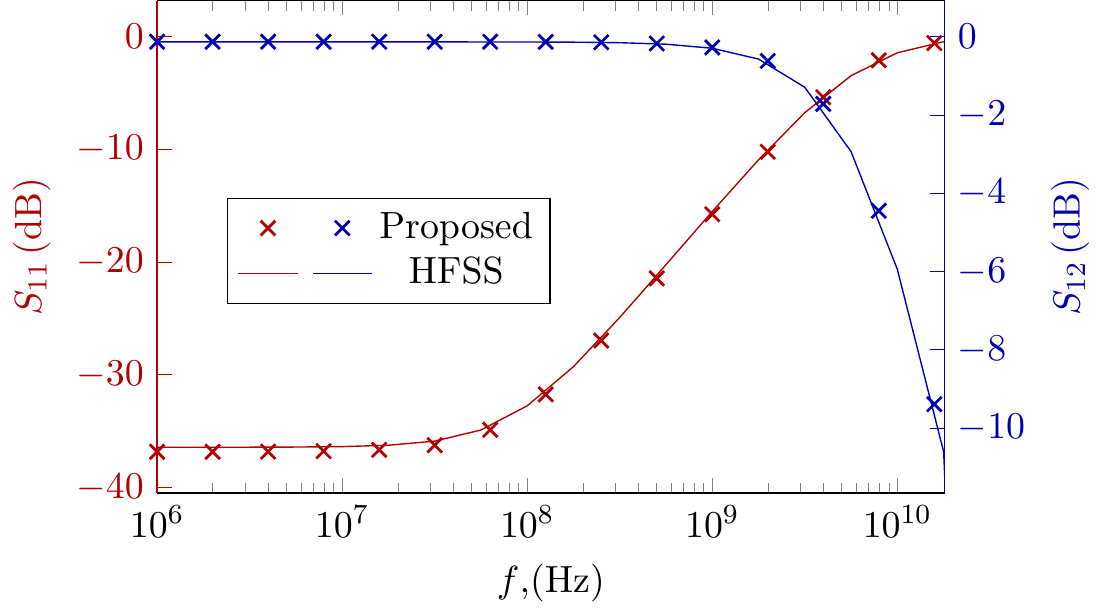}
		\label{inductor_multilayer_Slog}
	}
	\caption{Selected transmission ($S_{11}$) and reflection ($S_{12}$) parameters for the two-port inductor coil in~\secref{sec:results:inductor}, with (a) validation against DCIM, and (b) broadband validation against HFSS.}
	\label{inductor_multilayer_S}
\end{figure}

Scattering parameters computed with the proposed MGF technique are validated against DCIM in \figref{inductor_multilayer_Slin}. Also shown are the $S$ parameters obtained when only the QS contribution of the MGF is used in the near-region. Although the QS contribution is sufficient at low frequencies when the electrical size of the structure is small, it is clearly insufficient over the frequency range of interest. While QS terms alone are unable to correctly identify the resonant frequency, the proposed method is as accurate as DCIM.
The proposed solver is further validated against a commercial finite element solver (Ansys HFSS 18.2) in \figref{inductor_multilayer_Slog}, and performs very well from the megahertz to the gigahertz range.

Profile data is shown in \tabref{stats}; it is clear that the proposed method significantly outperforms HFSS in terms of time and memory.
The proposed MGF technique is also faster than DCIM in computing near-region matrix entries by a factor of $2$, but the overall savings are modest since the problem is quite small.

%
%

%
%

\subsection{On-chip Interconnect Network} \label{sec:results:interconnect}
\begin{figure}[t]
	\includegraphics[width=9cm, page=1, trim={0 0 0 22cm}, clip]{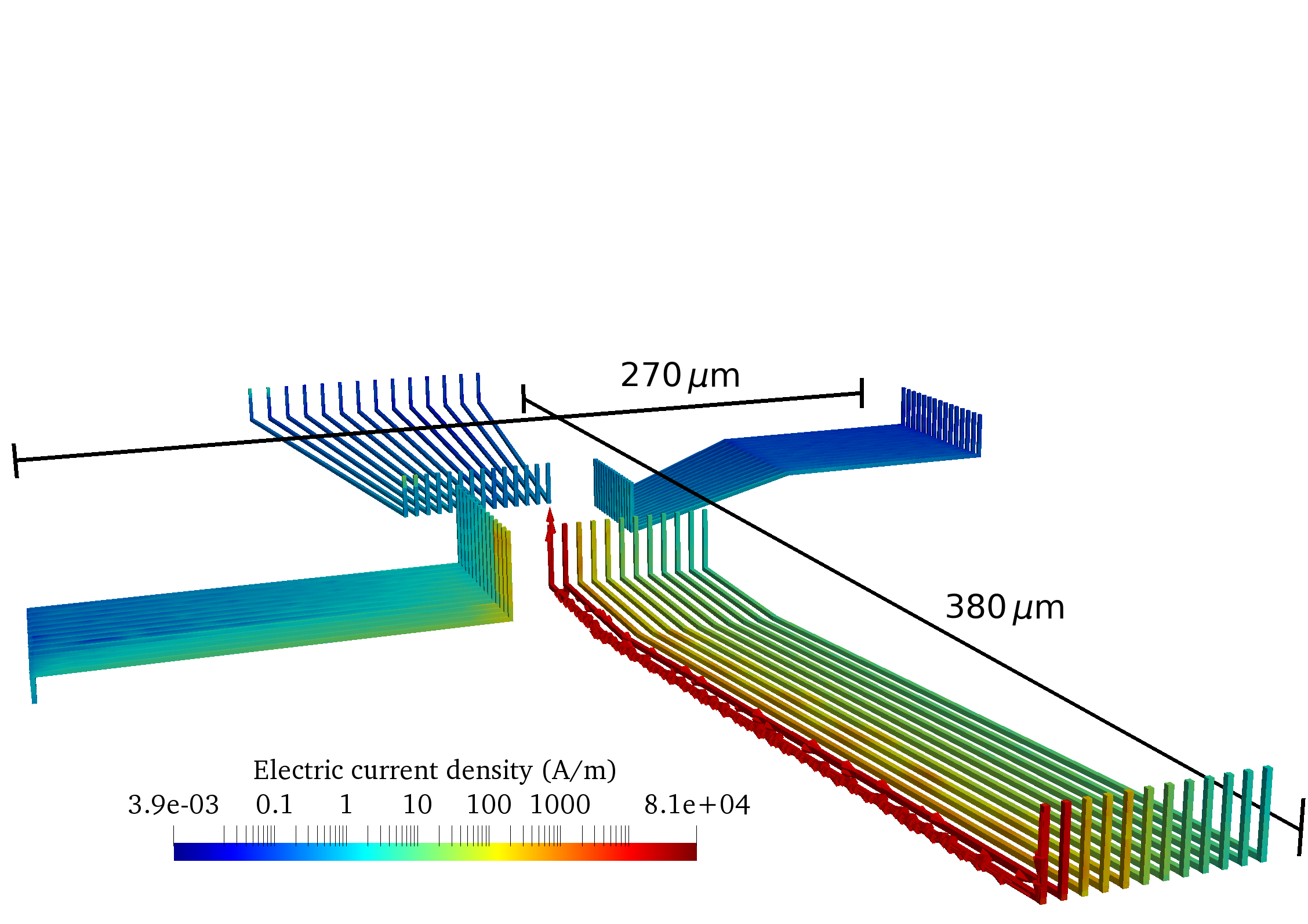}
	\caption{Computed surface current density distribution at $10$\,GHz on the interconnect network in~\secref{sec:results:interconnect}.}
	\label{toy_int_Js}
\end{figure}

\begin{figure}[t]
	\includegraphics[width=9cm, page=1, trim={0 0 0 0}, clip]{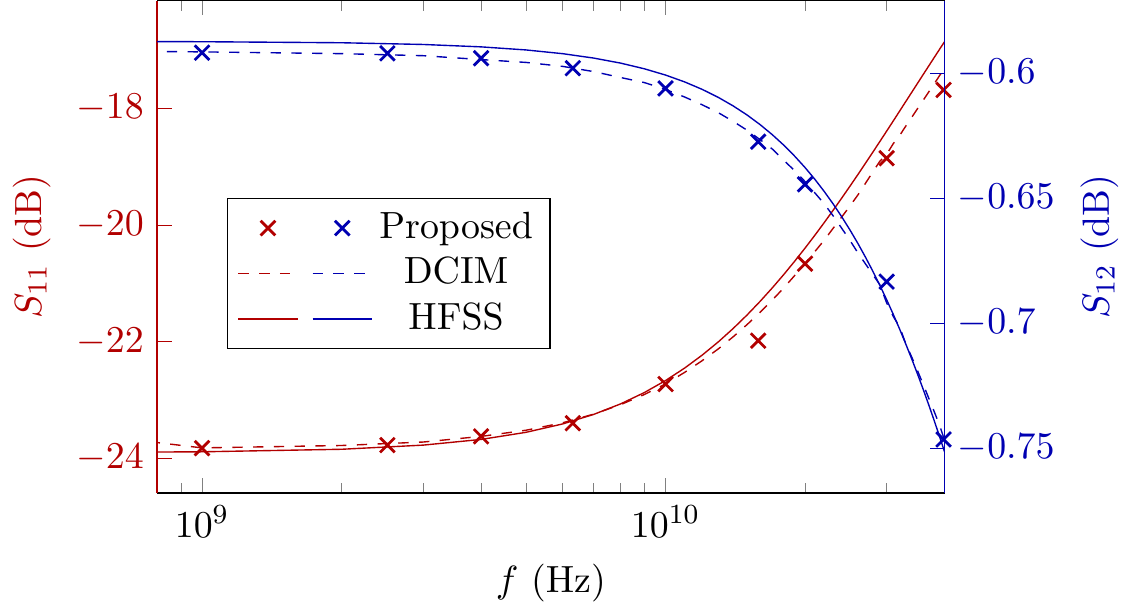}
	\caption{Selected transmission ($S_{11}$) and reflection ($S_{12}$) parameters for the four-port on-chip interconnect in~\secref{sec:results:interconnect}, compared with HFSS.}
	\label{toy_interposer}
\end{figure}

Next, we consider a four-port network of $55$ copper interconnects with a cross section of $1\,\mu$m$ \times 1\,\mu$m and conductor lengths of $150$\,$\mu$m on average. The geometry and current distribution are shown in \figref{toy_int_Js}. The interconnect network is embedded in five lossy dielectric layers whose configuration and properties are given in \tabref{Jnumcases}.
The layers are bounded by free space above and below. The $S$ parameters are validated against HFSS, as shown in \figref{toy_interposer}, and are in very good agreement.
Profile data in \tabref{stats} confirms performance benefits of the proposed MGF computation technique. A savings of over $2\times$ is obtained in near-region matrix CPU time, as compared to DCIM.
This has a significant impact on the overall solution time.
In the case of DCIM, $54\%$ of CPU time is spent on near-region computations.
With the proposed method, only $34\%$ of the total time is occupied by near-region computations, thus mitigating a major computational bottleneck.

\begin{table}[t]
	\centering
	\caption{CPU time for FFT-related computations for~the~interconnect~network~in~\secref{sec:results:interconnect}.}
	\begin{tabular}{lllc}
		\toprule
		& Multilayer & Homogeneous \\
		& (2D FFTs) & (3D FFT) \\
		\midrule
		Precorrection time (s) & 44 & 52 \\
		Average time per iteration (s) & 1.88 & 2.55 \\
		\bottomrule
	\end{tabular}
	\label{AIMstats}
\end{table}

\begin{figure}[t]
	\includegraphics[width=9cm, page=1, trim={0cm 0 0 0}, clip]{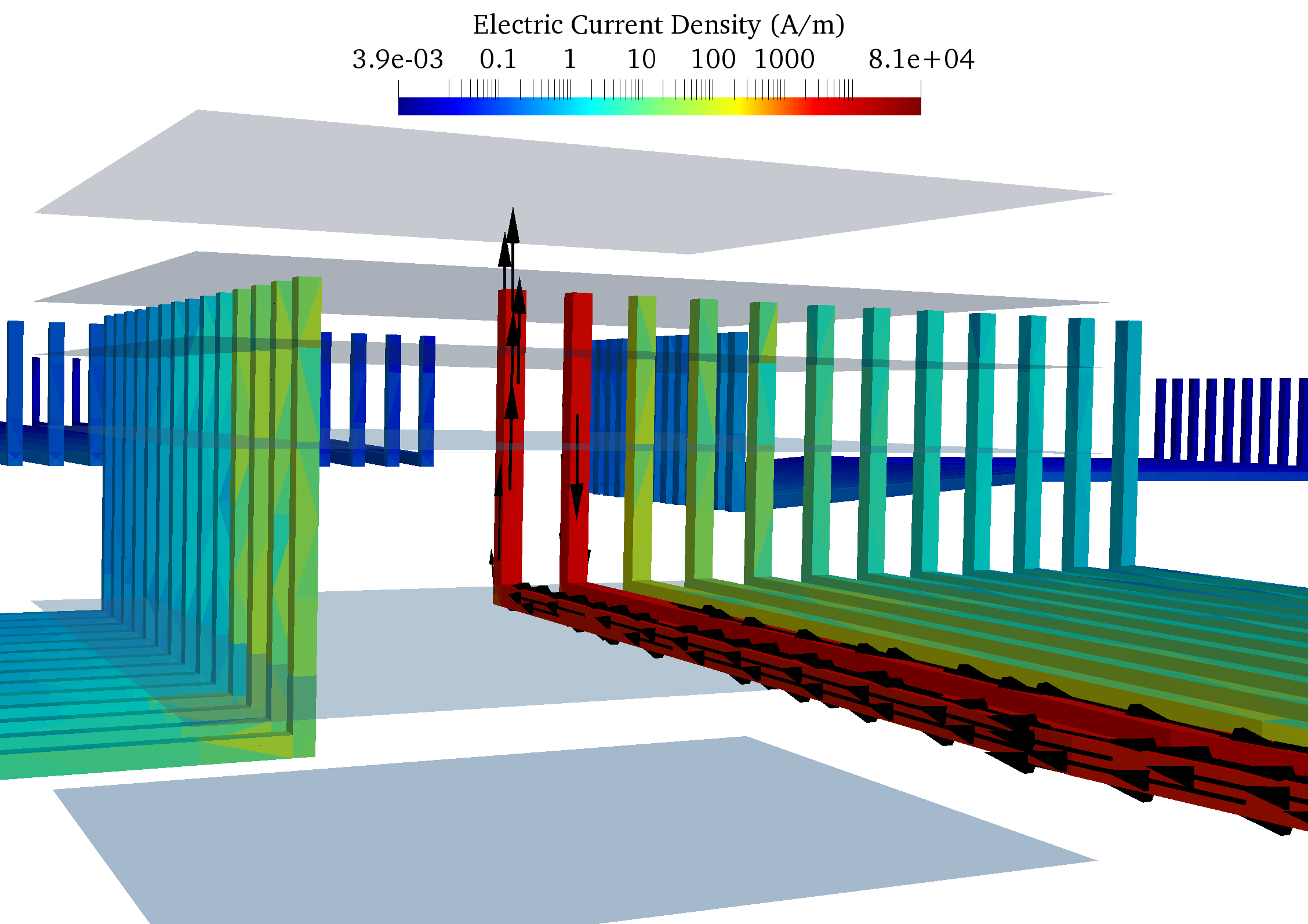}
	\caption{Validation of current continuity for conductors traversing multiple layers at $f = 10\,$GHz, for the interconnect network in \secref{sec:results:interconnect}.}
	\label{continuity_interconnect}
\end{figure}

To ensure that the proposed formulation is applicable to conductors traversing multiple layers, the surface current density on vias in the interconnect network is shown more closely in \figref{continuity_interconnect}. The interfaces between layers are also visualized to indicate points at which the vias are split. It is clear that the vectorial surface current density in both vertical and horizontal directions is smooth and well-behaved, confirming that continuity across layer interfaces is correctly enforced.

In order to confirm that the proposed 2D FFT-based modification to AIM is computationally comparable to conventional homogeneous AIM, we simulate the interconnect network in free space with 3D FFT-based AIM. FFT computation is encountered in the precorrection phase of AIM, and in each iteration of the matrix solve. Thus we compare the precorrection and average iteration time, for the 2D FFT (with MGF) and 3D FFT (in free space) approaches. The results are given in \tabref{AIMstats}, and confirm that the efficiency of the proposed MGF-based AIM is comparable to the homogeneous case.

%
%

\section{Conclusions} \label{sec:concl}
A complete surface integral solver is proposed to efficiently and accurately model lossy conductors in stratified dielectric media. A novel series expansion-based technique is described to accelerate the computation of the multilayer Green's function in the near-region, in the context of the adaptive integral method (AIM). A modification to AIM is proposed to handle 3D conductors in stratified media, whose efficiency is comparable to the homogeneous case. A differential surface admittance operator is used to capture skin effect inside conductors, and is generalized to multi-conductor systems embedded in layered media in any configuration, including conductors that traverse dielectric layers. Further, the augmented EFIE formulation is used to separate current and charge densities and obtain robust performance over a wide frequency range. The accuracy and speed of the solver are demonstrated on two realistic structures, including a large interconnect network. Scattering parameters are accurately computed over the entire frequency range of interest, and are validated against a commercial finite element tool. In each test case, the proposed MGF computation technique is over twice as fast as DCIM in computing near-region matrix entries.

\appendix
\subsection{Port Handling} \label{app:ports}
For interconnect problems, it is necessary to be able to extract scattering parameters of the network, given a set of terminals and ports. This requires coupling the system \eqref{systemEM} to excitation and load circuit elements.

We define subsets of mesh triangles $\{T_t\}_i$ as belonging to port terminals, where the index $i = 1, 2\ldots N_\mathrm{term}$ and $N_\mathrm{term}$ is the number of terminals. The index $t = 1, 2\ldots N_{t,i}$ where $N_{t,i}$ is the number of triangles in terminal $i$. Each port is then naturally defined as a set of two terminals. Coupling to terminals is achieved by augmenting~\eqref{systemEM} as~\cite{gope},
\begin{equation}
	\setlength\arraycolsep{1.7pt}
	\begin{bmatrix}
		\matr{Z}_{\mathrm{EM}} & -\matr{D}^T\matr{Z}_\Phi & \matr{0} & \matr{0} \\ 
		\matr{D} & jk_0\matr{I} & jk_0\matr{Z}_{\mathrm{JK}} & \matr{0} \\ 
		\matr{0} & \matr{Z}_{\mathrm{KJ}} & \matr{0} & \eta_{0}^{-1}\matr{C} \\ 
		\matr{0} & \matr{0} & \matr{R} & \matr{P} 
	\end{bmatrix}
	\begin{bmatrix}
		\matr{J}_\mathrm{eq} \\ c_0\matr{\rho} \\ \matr{J}_i \\ \matr{V}_t 
	\end{bmatrix} = 
	\begin{bmatrix}
		\matr{0} \\ \matr{0} \\ \matr{0} \\ \matr{V}_s 
	\end{bmatrix}.\label{system_noneutr}
\end{equation}
The additional equations in \eqref{system_noneutr} relate $J_i(\vecprime{r})$ to a Th\'evenin equivalent circuit consisting of a voltage source, $V_s$, and a series resistance $R$.
This is achieved as follows:
\begin{itemize}
	\item The continuity equation, which is the second row in~\eqref{system_noneutr}, is modified for terminal triangles,
	\begin{equation}
		\divg\vect{J}_\mathrm{eq}(\vecprime{r}) + j\omega\rho_s(\vecprime{r}) = J_i(\vecprime{r}). \label{continuity}
	\end{equation}
	Term $J_i$ is the volume current density injected into the system from an external circuit, and is discretized on mesh triangles using area-normalized pulse basis functions. The additional vector of unknowns $\matr{J}_i$ collects coefficients of the pulse basis expansion of $J_i(\vecprime{r})$. Matrix $\matr{Z}_\mathrm{JK}$ consists of ones in rows that correspond to terminal triangles, and zeros otherwise. Its purpose is to pick out and modify the continuity equations only for those triangles that are part of terminals.
	\item Terminal voltages are expressed in terms of scalar potentials,
	\begin{equation}
	V_t = \dfrac{1}{\epsilon_0}\int_{S'}{G_\phi(\vect{r},\vecprime{r})\,\rho_s(\vecprime{r})\,dS'}, \label{Vt}
	\end{equation}
	where $V_t$ is the voltage at terminal $t$. Assigning potentials to each terminal triangle, followed by expanding and testing with pulse basis functions leads to the third equation in~\eqref{system_noneutr}, where terminal voltage coefficients are stored in $\matr{V}_t$. Matrix $\matr{Z}_\mathrm{KJ}$ consists of scalar potentials on terminal triangles. Matrix $\matr{C}$ consists of ones and zeros to enforce a constant scalar potential over all triangles that constitute a single terminal.

	\item The fourth row in~\eqref{system_noneutr} is obtained by applying Kirchoff's voltage law to relate $J_i$ and $V_t$. A Th\'{e}venin-equivalent model for the external circuit is assumed between each pair of terminals that forms a port, with a source voltage $V_s$ and series resistance $R$. For a given port $p$ with input terminal $t_i$ and output terminal $t_o$, this yields
	\begin{equation}
		V_{t_i} - V_{t_o} = V_s + I_tR, \label{KVL}
	\end{equation}
	where $I_t$ is the current injected into the system by the external circuit. Since the volume current density $J_i$ is discretized with area-normalized pulse basis functions defined on terminal triangles, its coefficients have units of amperes, and correspond directly to the injected circuit current $I_t$.
	In~\eqref{system_noneutr}, matrix $\matr{P}$ contains the coefficients of $V_{t_i}$ and $V_{t_o}$, while $\matr{R}$ contains resistances $R$. Vector $\matr{V}_s$ stores source voltages at each terminal.
\end{itemize}

\subsection{Charge Neutrality} \label{app:charge}
The system in~\eqref{system_noneutr} does not account for charge neutrality on conductors, which leads to loss of rank at low frequencies. As suggested by Qian and Chew~\cite{chew1}, we handle this by dropping one charge density unknown for each unconnected conductor in the structure. To account for conductors connected to each other via ports, or otherwise in contact with each other, an adjacency matrix is constructed to find each set of conductors that is isolated from the others. One charge density unknown is then dropped for each set of connected conductors, rather than each individual conductor. Mathematically, this is achieved by introducing mapping matrices $\matr{F}$ and $\matr{B}$, defined previously~\cite{chew1}. These matrices map the full set of charge unknowns $\matr{\rho}$ to and from a reduced set of unknowns, $\matr{\rho}_r$. These matrices are incorporated into~\eqref{system_noneutr} to yield the final system,
\begin{equation}
	\setlength\arraycolsep{1.7pt}
	\begin{bmatrix}
		\matr{Z}_{\mathrm{EM}} & -\matr{D}^T\matr{Z}_\Phi\matr{B} & \matr{0} & \matr{0} \\ 
		\matr{F}\matr{D} & jk_0\matr{I}_r & \matr{F}\matr{Z}_{\mathrm{JK}} & \matr{0} \\ 
		\matr{0} & \matr{Z}_{\mathrm{KJ}}\matr{B} & \matr{0} & \eta_{0}^{-1}\matr{C} \\ 
		\matr{0} & \matr{0} & \matr{R} & \matr{P} 
	\end{bmatrix}
	\begin{bmatrix}
		\matr{J}_\mathrm{eq} \\ c_0\matr{\rho}_r \\ \matr{J}_i \\ \matr{V}_t 
	\end{bmatrix} = 
	\begin{bmatrix}
		\matr{0} \\ \matr{0} \\ \matr{0} \\ \matr{V}_s 
	\end{bmatrix}, \label{aEFIE-J}
\end{equation}
where $\matr{I}_r$ is the identity matrix of reduced size, corresponding to $\matr{\rho}_r$.

\subsection{Preconditioning} \label{app:PC}
In order to speed up convergence of the iterative matrix solution of~\eqref{aEFIE-J}, a good preconditioner is necessary. We employ a sparse right-preconditioner similar to the one in~\cite{aefie},
\begin{equation}
	\setlength\arraycolsep{1.7pt}
	\matr{M} = 
	\begin{bmatrix}
		\mathrm{diag}\left(\matr{Z}_{\mathrm{EM}}\right) & \mathrm{diag}\left(-\matr{D}^T\matr{Z}_\Phi\matr{B}\right) & \matr{0} & \matr{0} \\ 
		\matr{F}\matr{D} & jk_0\matr{I}_r & \matr{F}\matr{Z}_{\mathrm{JK}} & \matr{0} \\ 
		\matr{0} & \matr{Z}_{\mathrm{KJ}}\matr{B} & \matr{0} & \eta_{0}^{-1}\matr{C} \\ 
		\matr{0} & \matr{0} & \matr{R} & \matr{P} 
	\end{bmatrix}, \label{PC}
\end{equation}
where $\mathrm{diag}(\cdot)$ represents diagonal terms of the corresponding matrix.
The preconditioner can be abbreviated as
\begin{equation}
	\matr{M} = 
	\begin{bmatrix}
		\matr{Z}_{\mathrm{EM,d}} & \matr{M}_{1,2} \\ 
		\matr{M}_{2,1} & \matr{M}_{2,2}
	\end{bmatrix} \label{PC_abbr}
\end{equation}
where $\matr{Z}_{\mathrm{EM,d}} = \mathrm{diag}\left(\matr{Z}_{\mathrm{EM}}\right)$, and
\begin{subequations}
	\begin{align}
		\matr{M}_{1,2} &= \begin{bmatrix} \mathrm{diag}\left(-\matr{D}^T\matr{Z}_\Phi\matr{B}\right) & \matr{0} & \matr{0} \end{bmatrix}, \\
		\matr{M}_{2,1} &= \begin{bmatrix} \matr{F}\matr{D} & \matr{0} & \matr{0} \end{bmatrix}^T, \\
		\matr{M}_{2,2} &= \begin{bmatrix} jk_0\matr{I}_r & \matr{F}\matr{Z}_{\mathrm{JK}} & \matr{0} \\
		\matr{Z}_{\mathrm{KJ}}\matr{B} & \matr{0} & \eta_{0}^{-1}\matr{C} \\ 
		\matr{0} & \matr{R} & \matr{P} \end{bmatrix}.
	\end{align}
\end{subequations}

Leveraging the Schur complement of $\matr{Z}_{\mathrm{EM,d}}$, the exact inverse of this preconditioner can be written as
\begin{multline}
	\bs{M}^{-1} = 
	\begin{bmatrix} 
		\matr{Z}_{\mathrm{EM,d}}^{-1} & \matr{0} \\
		\matr{0} & \matr{0}
	\end{bmatrix} +\\
	\begin{bmatrix} 
		-\matr{Z}_{\mathrm{EM,d}}^{-1}\matr{M}_{1,2} \\
		\matr{I}
	\end{bmatrix}
	\matr{\Delta}^{-1}
	\begin{bmatrix} 
		-\matr{M}_{2,1}\matr{Z}_{\mathrm{EM,d}}^{-1} & \matr{I}
	\end{bmatrix} \label{PC_inv}
\end{multline}
where $\matr{\Delta} = \matr{M}_{2,2} - \matr{M}_{2,1}\matr{Z}_{\mathrm{EM,d}}^{-1}\matr{M}_{1,2}$. Since $\matr{Z}_{\mathrm{EM,d}}$ is diagonal, computing its inverse is trivial. The preconditioner is applied to the system matrix blocks during the matrix-vector product at each iteration.
The restarted-GMRES linear solver~\cite{gmres} provided through the PETSc software package~\cite{petsc-web-page} is used to iteratively solve~\eqref{aEFIE-J}.

\bibliographystyle{ieeetr}
\bibliography{inc/IEEEabrv,inc/biblio3D_Utkarsh,inc/biblio_Shashwat}

\end{document}